\documentclass[twocolumn,prb]{revtex4-1}

\usepackage{physics}
\usepackage[dvipdfmx]{graphicx}
\usepackage{amssymb}
\usepackage{amsmath}
\usepackage{multirow}
\usepackage{bm}
\usepackage{color}
\usepackage{mypack}
\usepackage{here}

\usepackage{comment}
\usepackage{color}
\usepackage[normalem]{ulem}
\usepackage{cancel}

\definecolor{rbrown}{rgb}{0.2,0.2,0.8}

\begin{document}

\title{Anisotropic superconducting spin transport at magnetic interfaces}
\author{Yuya Ominato${}^1$, Ai Yamakage${}^2$, and Mamoru Matsuo${}^{1,3,4,5}$}
\affiliation{${}^1$Kavli Institute for Theoretical Sciences, University of Chinese Academy of Sciences, Beijing 100190, China}
\affiliation{${}^2$Department of Physics, Nagoya University, Nagoya 464-8602, Japan}
\affiliation{${}^3$CAS Center for Excellence in Topological Quantum Computation, University of Chinese Academy of Sciences, Beijing 100190, China}
\affiliation{${}^4$Advanced Science Research Center, Japan Atomic Energy Agency, Tokai, 319-1195, Japan}
\affiliation{${}^5$RIKEN Center for Emergent Matter Science (CEMS), Wako, Saitama 351-0198, Japan}
\date{\today}

\begin{abstract}
We present a theoretical investigation of anisotropic superconducting spin transport at a magnetic interface between a $p$-wave superconductor and a ferromagnetic insulator. Our formulation describes the ferromagnetic resonance modulations due to spin current generation depending on spin-triplet Cooper pair, including the frequency shift and enhanced Gilbert damping, in a unified manner. We find that the Cooper pair symmetry is detectable from the qualitative behavior of the ferromagnetic resonance modulation. Our theory paves the way toward anisotropic superconducting spintronics.
\end{abstract}
\maketitle



{\it Introduction.---}Use of spin-triplet Cooper pairs as carriers for spin currents in the emergent field of superconducting spintronics is challenging \cite{linderSuperconductingSpintronics2015,hanSpinCurrentProbe2020}.
Previous studies have demonstrated spin transport mediated by spin-triplet Cooper pairs that formed at the $s$-wave superconductor (SC)/ferromagnet interfaces of Josephson junctions.
The spin-singlet pairs in SCs are converted into spin-triplet pairs in half-metallic $\mathrm{CrO}_2$ \cite{keizerSpinTripletSupercurrent2006}.
However, previous studies on spin-triplet pairs at magnetic interfaces have been limited to cases induced by the proximity effect.

One promising candidate material system for investigation of spin-triplet currents to enable more active use of spin-triplet pairs is the $p$-wave SC/ferromagnetic insulator (FI) bilayer thin film system \cite{tanakaTheoryTopologicalSpin2009,brydonSpontaneousSpinCurrent2009}.
Tunneling of the spins is driven by the magnetization dynamics excited by ferromagnetic resonance (FMR) in the ferromagnetic material via interfacial exchange coupling between the magnetization in the FI and the electron spins in the $p$-wave SC, and a spin-triplet current is expected to be generated.
Furthermore, as a backaction of spin injection, both the FMR frequency and the Gilbert damping of the FI should be modulated \cite{tserkovnyakEnhancedGilbertDamping2002,tserkovnyakNonlocalMagnetizationDynamics2005a,simanekGilbertDampingMagnetic2003}.
Although similar scenarios have already been studied vigorously in $s$-wave SC/ferromagnet systems, most previous studies have focused on the Gilbert damping modulation due to spin injection \cite{inoueSpinPumpingSuperconductors2017,katoMicroscopicTheorySpin2019,silaevFinitefrequencySpinSusceptibility2020,silaevLargeEnhancementSpin2020,ojajarviNonlinearSpinTorque2020,simensen2021spin,bellSpinDynamicsSuperconductorFerromagnet2008,jeonEnhancedSpinPumping2018,yaoProbeSpinDynamics2018,liPossibleEvidenceSpinTransfer2018,jeonEffectMeissnerScreening2019,jeonEnhancedSpinPumping2018,jeonAbrikosovVortexNucleation2019,golovchanskiyMagnetizationDynamicsProximityCoupled2020,zhaoExploringContributionTrapped2020}. 
To gain an in-depth understanding of the spin-triplet transport mechanisms, the FMR modulation processes, including both the frequency shift and the enhanced Gilbert damping, should be formulated microscopically in a systematic manner.

Determination of the pairing symmetry of the spin-triplet $p$-wave SCs within the same framework is also desirable.
Despite many years of research based on several experimental techniques that detect the pairing symmetry, including nuclear magnetic resonance \cite{leggettTheoreticalDescriptionNew1975}, polarized neutron scattering \cite{shullElectronicNuclearPolarization1963,shullNeutronSpinNeutronOrbit1963,shullNeutronDiffractionStudiesElectronSpin1966}, and muon-spin resonance techniques \cite{lukeMuonSpinRelaxation1993}, there are few established candidate systems for spin-triplet SCs \cite{saxenaSuperconductivityBorderItinerantelectron2000,aokiCoexistenceSuperconductivityFerromagnetism2001,huySuperconductivityBorderWeak2007,ranNearlyFerromagneticSpintriplet2019a,yang2021spin}.
The FMR modulation has been observed in various nanoscale magnetic multilayers. Accordingly, the technique is widely used to investigate a spin transport property in a variety of nanoscale thin film systems because it is highly sensitive.
Thus one can expect that the FMR measurements in $p$-wave SC/FI bilayer systems provide useful information about pairing symmetry.

\begin{figure}[t]
\begin{center}
\includegraphics[width=1\hsize]{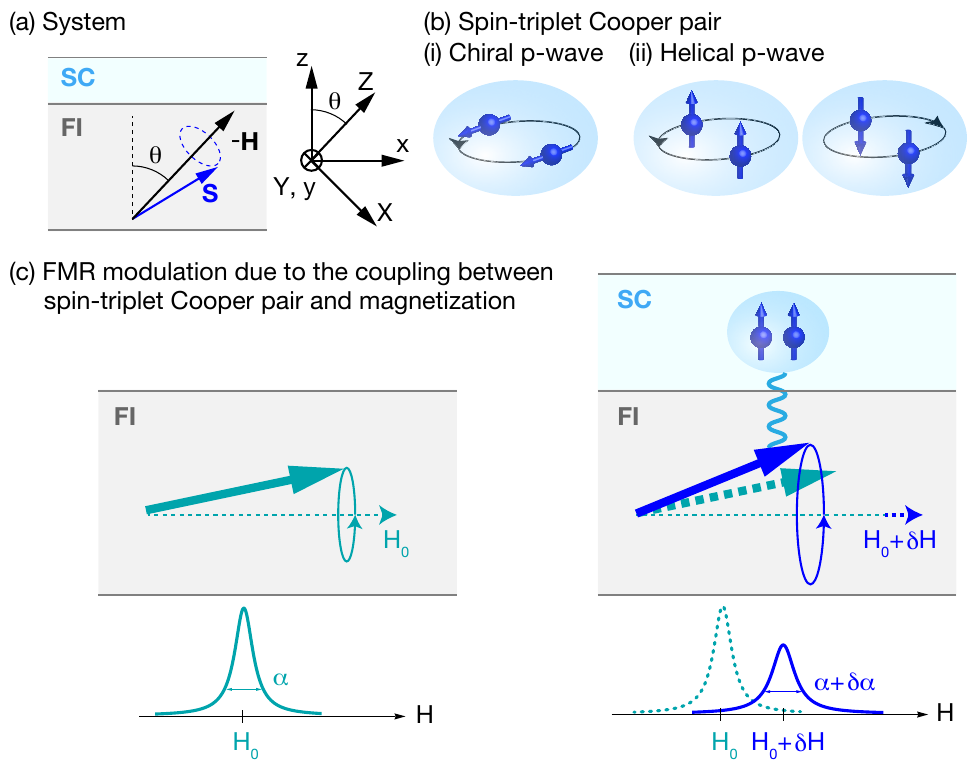}
\end{center}
\caption{Mechanism of FMR modulation due to anisotropic superconducting spin transport at magnetic interfaces.
(a) Precession axis located on the $x$-$z$ plane, where the angle between the precession axis and the $z$ axis is $\theta$ (where $0\leq\theta\leq\pi/2$).
(b) Two types of spin-triplet Cooper pairs considered in this work.
(c) FMR signal modulation in the SC/FI bilayer system compared with the signal in the FI monolayer.
}
\label{fig_system}
\end{figure}

In this Letter, we investigate anisotropic superconducting spin transport at the magnetic interfaces of hybrid systems composed of $p$-wave SC/FI thin films theoretically, as illustrated in Fig.\ \ref{fig_system}(a).
The two-dimensional bulk SC is placed on the FI, where the FMR occurs.
The precession axis is rotated by an angle $\theta$ from the direction perpendicular to the interface.
Here, we use two coordinate systems: $(x,y,z)$ and $(X,Y,Z)$. The $z$ axis is perpendicular to the interface and the $x$ and $y$ axes are along the interface.
The $(X,Y,Z)$ coordinate is obtained by rotating the angle $\theta$ around the $y$ axis, so that the precession axis and the $Z$ axis are parallel.
Figure \ref{fig_system}(b) shows a schematic image of the spin-triplet Cooper pairs for the chiral and helical $p$-wave SCs considered in this work.
Figure \ref{fig_system}(c) shows a schematic image of the FMR signal in the FI monolayer and the SC/FI bilayer.
The FMR frequency and linewidth in the SC/FI bilayer are both modulated because of the spin transfer occurring at the interface.

Using the nonequilibrium Green's function method, we formulate the FMR modulations due to the back action of the spin-triplet transport process systematically.
The main advantage of using the nonequilibrium Green's function is dealing with both a spectral function and a nonequilibrium distribution function.
Indeed, the interface spin current is given by the expression using the nonequilibrium distribution function, which shows that the interface spin current by the spin pumping and the enhanced Gilbert damping are proportional to each other.
Furthermore, as an advantage of field theoretical treatment, the frequency shift and the enhanced Gilbert damping are both described in a unified manner.
Additionally, it is shown that the symmetry of the spin-triplet pairs can be extracted from the FMR modulations.
The results presented here offer a pathway toward development of anisotropic superconducting spintronics.

{\it Model Hamiltonian.---}The FMR modulation due to the SC adjacent to the FI is calculated microscopically using the spin tunneling Hamiltonian method \cite{ohnumaEnhancedDcSpin2014,ohnumaTheorySpinPeltier2017,tataraConsistentMicroscopicAnalysis2017,matsuoSpinCurrentNoise2018,ominatoQuantumOscillationsGilbert2020a,ominatoValleyDependentSpinTransport2020a,inoueSpinPumpingSuperconductors2017,katoMicroscopicTheorySpin2019,silaevFinitefrequencySpinSusceptibility2020}. The effect of the SC on the FI is treated as a perturbation and suppression of ferromagnetism with the onset of superconductivity is assumed to be negligible, which is consistent with the results of spin pumping experiments in magnetic multilayer thin films. The details of the model Hamiltonians and the formulations are described in the Supplemental Material \cite{sm}.
In the main text, we focus on giving an overview of the model Hamiltonians and the formulations.

The total Hamiltonian $H(t)$ comprises three terms
\begin{align}
    H(t)=\hfi(t)+\hsc+\hex.
\end{align}
The first term $\hfi(t)$ describes the bulk FI,
\begin{align}
\hfi(t)=
    &\sum_\bk\ho_\bk \bc\ba
    -\hac^+(t)b^\dagger_{\bk=\bm{0}}
    -\hac^-(t)b_{\bk=\bm{0}},
\end{align}
where $\bc$ and $\ba$ denote the creation and annihilation operators of magnons with the wave vector $\bk=(k_x,k_y,k_z)$, respectively.
We assume the parabolic dispersion $\hbar\omega_\bk=Dk^2-\hbar\gamma H$, where $\gamma(<0)$ is the electron gyromagnetic ratio.
The coupling between the microwave radiation and the magnons is given by
$\hac^\pm(t)=\hbar\gamma\hac\sqrt{SN/2}e^{\mp i\omega t}$,
where $\hac$ and $\omega$ are the amplitude and the frequency of the microwave radiation, respectively.
$S$ is the magnitude of the localized spin and $N$ is the number of sites in the FI.
Note that the precession axis for the localized spin is fixed along the $Z$ axis [see Fig.\ \ref{fig_system}(a)].

The second term $H_{\rm SC}$ describes the two-dimensional bulk SCs,
\begin{align}
    H_{\mathrm{SC}}=
        \frac{1}{2}\sum_{\bk}
        \bm{c}_{\bk}^{\dagger}
        H_{\mathrm{BdG}}\bm{c}_{\bk},
\end{align}
where we use the four-component notations
\begin{align}
    &\bm{c}_{\bk}^\dagger
    =
    (
        c_{\bk\uparrow}^\dagger,
        c_{\bk\downarrow}^\dagger,
        c_{-\bk\uparrow},
        c_{-\bk\downarrow}
    ), \\
    &\bm{c}_{\bk}
    =
    (
        c_{\bk\uparrow},
        c_{\bk\downarrow},
        c_{-\bk\uparrow}^\dagger,
        c_{-\bk\downarrow}^\dagger
    )^{\rm T}.
\end{align}
Here, $c_{\bk s}^\dagger$ and $c_{\bk s}$ denote creation and annihilation operators, respectively, of electrons with the wave vector $\bk=(k_x,k_y)$ and the $z$ component of the spin $s=\uparrow,\downarrow$.
The Bogoliubov-de Gennes Hamiltonian $H_{\rm BdG}$ is a $4\times 4$ matrix given by
\begin{align}
    H_{\rm BdG}=
        \begin{pmatrix}
             \xi_{\bk}\sigma^0   & \Delta_\bk \\
            -\Delta^{\ast}_{-\bk} & -\xi_{\bk}\sigma^0
        \end{pmatrix},
\end{align}
where $\xi_\bk$ represents the energy of the electrons as measured from their chemical potential, $\sigma^0$ is a $2\times 2$ unit matrix, and the pairing potential $\Delta_{\bk}$ is also a $2\times 2$ matrix.
We consider three pairing potential types, including the spin-singlet $s$-wave pairing $\Delta_{\bk}=\Delta i\sigma^y$ and two spin-triplet $p$-wave pairings $\Delta_{\bk}=(\bm{d}_\bk\cdot\bm{\sigma})i\sigma^y$, where their $d$ vectors are given by
\begin{align}
    \bm{d}_{\bk}=
        \begin{cases}
            \Delta(0,0,e^{i\phi_\bk})            & :{\rm Chiral} {~}p{\rm -wave} \\
            \Delta(-\sin\phi_\bk,\cos\phi_\bk,0) & :{\rm Helical}{~}p{\rm -wave}
        \end{cases}
\end{align}
where $\phi_\bk=\arctan(k_y/k_x)$ is an azimuth angle.
The phenomenological form of the gap function is assumed
\begin{align}
    \Delta=1.76k_{\mathrm{B}}T_c\tanh\left(1.74\sqrt{T_c/T-1}\right),
\end{align}
with $T_c$ the superconducting transition temperature.
By diagonalizing $H_{\mathrm{BdG}}$, the quasiparticle energy is given by $E_{\bk}=\sqrt{\xi_\bk^2+\Delta^2}$ for all SCs considered here. Therefore, one cannot distinguish them by the energy spectrum alone, and they are simple models suitable for studying the difference of the magnetic responses due to the pairing symmetry \cite{sigristPhenomenologicalTheoryUnconventional1991}.

The third term $\hex$ represents the proximity exchange coupling that occurs at the interface, which describes the spin transfer between the SC and the FI \cite{ohnumaEnhancedDcSpin2014,katoMicroscopicTheorySpin2019},
\begin{align}
    &\hex
        =\sum_{\bq,\bk}
            \left(
                J_{\bq,\bk}
                \sigma^+_\bq
                S^-_\bk
                +
                \mathrm{h.c.}
            \right),
\end{align}
where $J_{\bq,\bk}$ is the matrix element for the spin transfer processes, $\sigma^\pm_{\bq}=(\sigma^X_{\bq} \pm i\sigma^Y_{\bq})/2$ represent the spin-flip operators for the electron spins in the SCs, and $S^-_{-\bk}=\sqrt{2S}b^\dagger_{\bk}$ and $S^+_{\bk}=\sqrt{2S}b_{\bk}$ represent the Fourier component of the localized spin in the FI.
Note that the precession axis is along the $Z$ axis, so that the $Z$ component of the spin is injected into the SC when the FMR occurs.
Using the creation and annihilation operators of electrons and magnons, $\hex$ is written as
\begin{align}
    \hex=\sum_{\bq,\bk,\bkp,s,s^\prime}
        \left(
            \sqrt{2S}J_{\bq,\bk}
            \sigma^+_{ss^\prime}c^\dagger_{\bkp s}c_{\bkp+\bq s^\prime}
            b_{-\bk}^\dagger
            +
            \mathrm{h.c.}
        \right).
\end{align}
From the above expression, one can see that $\hex$ describes electron scattering processes with magnon emission and absorption.

{\it Modulation of FMR.---}The FMR modulation can be read from the retarded component of the magnon Green's function \cite{ohnumaEnhancedDcSpin2014}, which is given by
\begin{align}
    &G_\bk^R(\omega)
        =\frac
            {2S/\hbar}
            {
                \omega
                -\omega_\bk
                +i\alpha\omega
                -(2S/\hbar)\Sigma^R_\bk(\omega)
            },
\end{align}
where the Gilbert damping constant $\alpha$ is introduced phenomenologically \cite{kasuyaRelaxationMechanismsFerromagnetic1961,cherepanovSagaYIGSpectra1993,jinTemperatureDependenceSpinwave2019}.
In the second-order perturbation calculation with respect to the matrix element $J_{\bq,\bk}$, the self-energy caused by proximity exchange coupling is given by
\begin{align}
    \Sigma^R_{\bk}(\omega)=-\sum_{\bq}|J_{\bq,\bk}|^2\chi_{\bq}^{R}(\omega),
    \label{eq_self_energy_J2chi}
\end{align}
where the dynamic spin susceptibility of the SCs is defined as
\begin{align}
    &\chi_{\bq}^{R}(\omega):=
        \int dte^{i(\omega+i0) t}
        \frac{i}{\hbar}
        \theta(t)
        \la[
            \sigma_{ \bq}^+(t),
            \sigma_{-\bq}^-(0)
        ]\ra.
\end{align}

The pole of $G_{\bk}^R(\omega)$ indicates the FMR modulation, i.e., the shift of resonance frequency and the enhancement of the Gilbert damping. By solving the equation
\begin{align}
    \omega-\omega_{\bk=\bm{0}}
    -(2S/\hbar)\mathrm{Re}\Sigma^R_{\bk=\bm{0}}(\omega)=0,
\end{align}
at a fixed microwave frequency $\omega$, one obtains the magnetic field at which the FMR occurs.
The imaginary part of the self-energy gives the enhancement of the Gilbert damping.
Consequently, the frequency shift and the enhanced Gilbert damping are given by
\begin{align}
    \delta H=
        \frac{2S}{\gamma\hbar}
        \mathrm{Re}\Sigma^R_{\bk=\bm{0}}(\omega),
        \hspace{2mm}
        \delta\alpha=
        -\frac{2S}{\hbar\omega}
        \mathrm{Im}\Sigma^R_{\bk=\bm{0}}(\omega).
        \label{eq_delta_H_delta_alpha}
\end{align}
From the above equations and Eq.~(\ref{eq_self_energy_J2chi}), one can see that the FMR modulation provides information about both the interface coupling properties and the dynamic spin susceptibility of the SCs.


The form of matrix element $J_{\bq,\bk=\bm{0}}$ depends on the details of the interface. In this work, we assume the interface with uncorrelated roughness. $|J_{\bq,\bk=\bm{0}}|^2$ is given by
\begin{align}
    |J_{\bq,\bk=\bm{0}}|^2
    =
    \frac{J_1^2}{N}\delta_{\bq,\bm{0}}
    +
    \frac{J_2^2l^2}{NA},
    \label{eq_matrix_element_model}
\end{align}
where the first and second terms describe averaged uniform contribution and uncorrelated roughness contribution, respectively \cite{sm}.
$J_1$ and $J_2$ correspond to the mean value and variance, respectively. $A$ is the area of the interface, which is equal to the system size of the SC. $l$ is an atomic scale length. Using Eq.\ (\ref{eq_matrix_element_model}), the self-energy for the uniform magnon mode is given by
\begin{align}
    &\Sigma^R_{\bk=\bm{0}}(\omega)
    =
    -\frac{J_1^2}{N}    \chi^R_{\mathrm{uni}}(\omega)
    -\frac{J_2^2l^2}{NA}\chi^R_{\mathrm{loc}}(\omega),
    \label{eq_self_energy_uni_loc}
\end{align}
where the uniform and local spin susceptibilities are defined as
\begin{align}
    \chi^{R}_{\mathrm{uni}}(\omega):=\lim_{|\bq|\to0}\chi^{R}_{\bq}(\omega),
    \hspace{2mm}
    \chi^R_{\mathrm{loc}}(\omega):=\sum_{\bq}\chi^R_{\bq}(\omega).
\end{align}
The self-energy $\Sigma^R_{\bk=\bm{0}}(\omega)$ consists of two terms originating from the uniform and roughness contributions, so that both $\chi^R_{\mathrm{uni}}(\omega)$ and $\chi^R_{\mathrm{loc}}(\omega)$ contribute to $\delta H$ and $\delta\alpha$.

Here, we discuss the FI thickness dependence on the FMR modulation \cite{chenMinimalModelSpinTransfer2015}.
From Eqs.~(\ref{eq_delta_H_delta_alpha}), and (\ref{eq_self_energy_uni_loc}), one can see that the FMR modulation is inversely proportional to the FI thickness ($\propto A/N$) because $\chi^R_{\mathrm{uni}}(\omega)\propto A$ and $\chi^R_{\mathrm{loc}}(\omega)\propto A^2$.
This is consistent with the experiments on the spin pumping in $\mathrm{Y}_3\mathrm{Fe}_5\mathrm{O}_{12}/\mathrm{Pt}$ heterostructures \cite{jungfleischThicknessPowerDependence2015}.
In order to observe the FMR modulation experimentally, it is necessary to prepare a sample  that is sufficiently thin, e.g., typically, the thickness of several tens of nanometers.

\begin{figure}[t]
\begin{center}
\includegraphics[width=1\hsize]{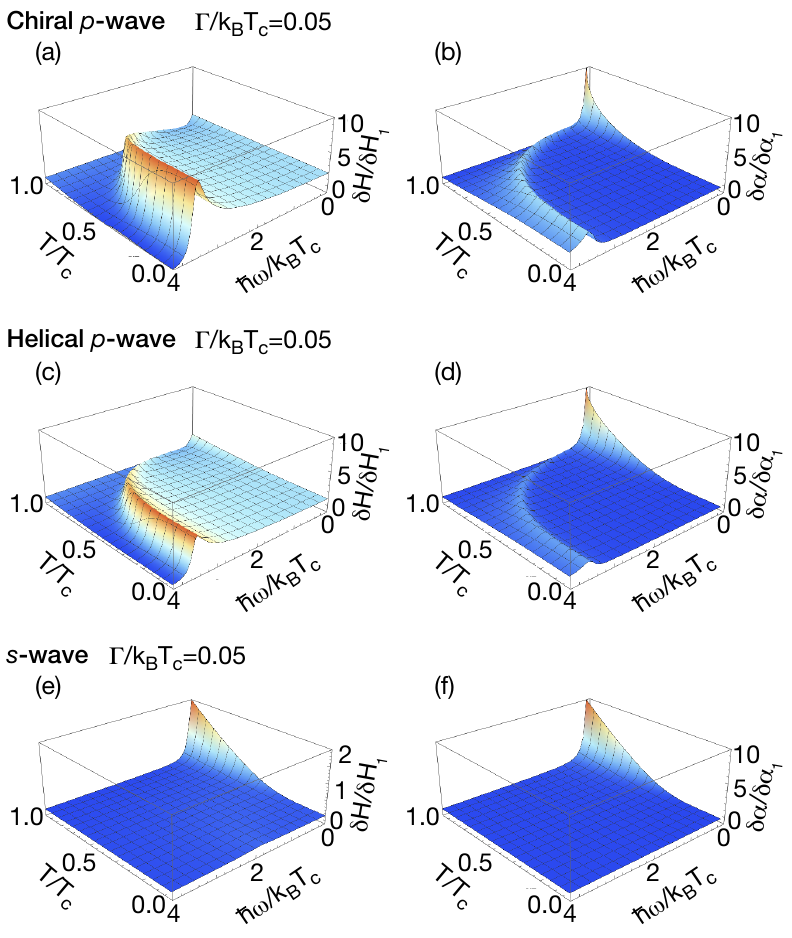}
\end{center}
\caption{The frequency shift $\delta H$ and the enhanced Gilbert damping $\delta\alpha$ as a function of temperature and frequency normalized by the characteristic values $\delta H_1 = -{SJ_1^2D_F}/({N\gamma\hbar})$ and $\delta\alpha_1 = {SJ_1^2D_F}/(Nk_\mathrm{B}T_c)$ in the normal state.
$D_F$ $(\propto A)$ is the density of states at the Fermi level in the normal state.
We set $\theta=0$ and $\Gamma/k_\mathrm{B}T_c=0.05$.
The sign of $\delta H$ corresponds to the sign of $\mathrm{Re}\chi^R_{\mathrm{uni}}(\omega)$, which can be positive and negative at low and high frequencies, respectively. In contrast, $\delta\alpha$ is positive at any frequency.}
\label{fig_chi_uni}
\end{figure}

{\it Numerical results.---} In the following, we consider a flat interface where $J_2=0$, so that the behavior of the FMR modulation is determined by $\chi^R_{\mathrm{uni}}(\omega)$.
The roughness contribution proportional to $\chi^R_{\mathrm{loc}}(\omega)$ is discussed later. Figure \ref{fig_chi_uni} shows the frequency shift $\delta H$ and the enhanced Gilbert damping $\delta\alpha$
as a function of temperature and frequency.
Here, we set $\theta=0$ and $\Gamma/k_\mathrm{B}T_c=0.05$, where $\Gamma$ is a constant level broadening of the quasiparticle introduced phenomenologically \cite{sm}.

First, we explain the qualitative properties of $\delta H$ and $\delta\alpha$ for the chiral $p$-wave SC.
In the low frequency region, where $\hbar\omega/k_{\mathrm{B}}T_c\leq 1$, $\delta H$ is finite and remains almost independent of $\omega$ near the zero temperature and $\delta\alpha$ decreases and becomes exponentially small with the decrease of the temperature.
In the high frequency region, where $\hbar\omega/k_{\mathrm{B}}T_c\geq 1$, a resonance peak occurs at $\hbar\omega=2\Delta$ for both $\delta H$ and $\delta\alpha$.
The qualitative properties of $\delta H$ and $\delta\alpha$ for the helical $p$-wave SC are the same as those of the chiral $p$-wave SC.

Next, we explain the qualitative properties of $\delta H$ and $\delta\alpha$ for the $s$-wave SC.
In the low frequency region, where $\hbar\omega/k_{\rm B}T_c\leq 1$, both $\delta H$ and $\delta\alpha$ decrease and become exponentially small with the decrease of the temperature.
In the high frequency region, where $\hbar\omega/k_{\rm B}T_c\geq 1$, both $\delta H$ and $\delta\alpha$ vanish.

The $p$-wave SCs show two characteristic properties that the $s$-wave SC does not show: a finite $\delta H$ at $T=0$ and a resonance peak of $\delta H$ and $\delta\alpha$.
These properties can be understood by the analogy between SCs and band insulators as follows.
The uniform dynamic spin susceptibility consists of contributions from intraband transitions within particle (hole) bands and interband transitions between particles and holes.
In the low temperature or high frequency region, the intraband contribution is negligible and the interband contribution is dominant.
In the case of the $s$-wave SC, the interband transitions are forbidden because 
the Hamiltonian and the spin operator commute. As a result, there is no spin response in the low-temperature or high-frequency regions.
In contrast, the Hamiltonian for the $p$-wave SCs and the spin operator do not commute.
Therefore, $\delta H$ has a finite value near-zero temperature due to the interband contribution. In addition, a resonance peak occurs when $\hbar\omega = 2 \Delta$ because the density of states diverges at the band edge $E = \pm\Delta$.
A detailed proof of the above statement is given in the Supplemental Material \cite{sm}.

\begin{figure}[t]
\begin{center}
\includegraphics[width=1\hsize]{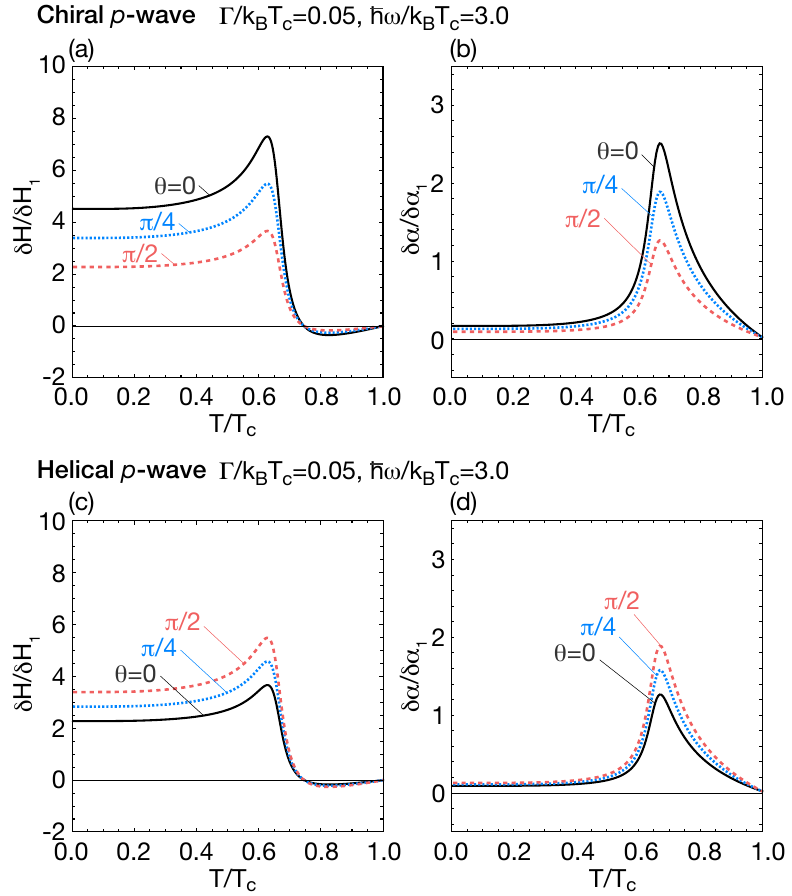}
\end{center}
\caption{
Frequency shift and the enhanced Gilbert damping as a function of temperature at angles of $\theta=0, \pi/4, \pi/2$.
The upper and lower panels show the characteristics for the chiral and helical $p$-wave SCs, respectively.
}
\label{fig_chi_uni_q}
\end{figure}

The angle dependences of $\delta H$ and $\delta\alpha$ are distinct for chiral and helical $p$-wave SCs, as shown in Fig.~\ref{fig_chi_uni_q}.
In both cases, we set $\hbar\omega/k_{\rm B}T_c=3.0$ as the typical values at high frequencies, where the main contribution of the uniform spin susceptibility is the interband transitions.
In the chiral $p$-wave SC, $\delta H$ and $\delta\alpha$ tend to decrease and are halved at a fixed temperature when $\theta$ increases from $0$ to $\pi/2$. Conversely, in the helical $p$-wave SC, the qualitative behavior shows the opposite trend. $\delta H$ and $\delta\alpha$ both tend to increase and become $1.5$ times larger at a fixed temperature when $\theta$ increases from $0$ to $\pi/2$.
In fact, the angle dependences are approximately obtained to be $\propto 1 + \cos^2\theta$ and $1 + (\sin^2\theta)/2$ for chiral and helical $p$-wave SCs, respectively \cite{sm}.
Therefore, the spin configuration of the Cooper pair can be detected from the $\theta$ dependence data for the FMR modulation.

The FMR modulation properties of the three SCs are summarized in Table \ref{table_chi_uni}.
All SCs considered here can be distinguished based on three properties: the frequency shift in the low temperature limit, the presence of their resonance peak, and their $\theta$ dependence.
For the $s$-wave SC, $\delta H$ becomes exponentially small in $T\to0$, while for the $p$-wave SCs, $\delta H$ is finite in $T\to0$.
For the $s$-wave SC, $\delta H$ and $\delta\alpha$ show no resonance and no $\theta$ dependence, while for the chiral and helical $p$-wave SCs, both $\delta H$ and $\delta\alpha$ exhibit a resonance at $\hbar\omega=2\Delta$ and a $\theta$ dependence. In addition, these two $p$-wave SCs can be distinguished from their $\theta$ dependences of $\delta H$ and $\delta\alpha$, which are characterized by $\partial_\theta(\delta H)$ and $\partial_\theta(\delta\alpha)$, respectively.
Here, it should be emphasized that the pairing symmetry can be characterized by the sign of $\partial_\theta(\delta H)$ and $\partial_\theta(\delta\alpha)$.
These properties are summarized in the Table \ref{table_chi_uni}.

\begin{table}[b]
\caption{\label{table_chi_uni}
FMR modulation properties for the flat SC/FI interface where $J_1\neq0$ and $J_2=0$.
}
\begin{ruledtabular}
\begin{tabular}{cccc}
    Pairing symmetry&
    $s$&
    Chiral&
    Helical\\
    \hline\hline
    $\delta H$ in the limit of $T\to0$ &
    $0$&
    finite &
    finite \\
    \hline
    Resonance peak of $\delta H$, $\delta\alpha$ &
    --&
    $\checkmark$ &
    $\checkmark$ \\
    \hline
    $\partial_\theta(\delta H)$, $\partial_\theta(\delta\alpha)$&
    $0$&
    negative&
    positive\\
\end{tabular}
\end{ruledtabular}
\end{table}

{\it Spin-triplet current generation.---}The relationship between the enhanced Gilbert damping discussed above and the spin-triplet current generation must also be discussed. 
The enhancement of the Gilbert damping is known to originate from the spin current generation at the magnetic interface \cite{tserkovnyakEnhancedGilbertDamping2002,ohnumaEnhancedDcSpin2014}.
The interface spin current induced by FMR $\la\IS\ra^{\mathrm{SP}}$ is given by \cite{sm}
\begin{align}
    \la\IS\ra^{\mathrm{SP}}
    =
    \frac{N(\hbar\gamma h_{\mathrm{ac}})^2}{2\alpha}
    \left[-{\mathrm{Im}}G^R_{\bk=\bm{0}}(\omega)\right]
    \delta\alpha.
\end{align}
One can see that $\la\IS\ra^{\mathrm{SP}}$ and $\delta\alpha$ are proportional  to each other.
In our setup, the enhanced Gilbert damping $\delta\alpha$ will lead to the generation of both the Cooper pair spin-triplet current and the quasiparticle spin current.
Since the angular dependence of $\delta\alpha$ reflects the direction of the Cooper pair spins, it is expected that the spin-triplet current can be controlled by varying the magnetization direction of the FI.

{\it Discussion.---}We have considered a flat SC/FI interface. 
In the presence of roughness, the correction term proportional to $\chi^R_{\mathrm{loc}}(\omega)$ contributes to the FMR modulation, as shown in Eq.~(\ref{eq_self_energy_uni_loc}).
In the rough limit, $J_1^2 \ll J_2^2$, $\chi^R_{\mathrm{loc}}(\omega)$ dominates to make the FMR modulation isotropic, due to the angle average by summation over $\bq$. 
Namely, the anisotropy peculiar to $p$-wave SC is smeared by the roughness. The detailed behavior of  $\chi^R_{\mathrm{loc}}(\omega)$ is shown in the Supplemental Material \cite{sm}.
This result implies that it is crucial to control the interface roughness.
In principle, the roughness of the interface can be observed using transmission electron microscopy of interfaces \cite{qiuSpinMixingConductance2013a,mihalceanuSpinpumpingVaryingthicknessMgO2017a,ikhtiarGiantTunnelMagnetoresistance2018a} and it is possible to detect whether the interface of the sample is flat or rough. More detailed spectroscopy can be obtained from the FMR modulation by using a flat interface.

Our results show that the pairing symmetry can be detected by the sign of $\partial_\theta(\delta H)$ and $\partial_\theta(\delta\alpha)$ around the in-plane magnetic field ($\theta \sim \pi/2$), where the vortices are negligible. When the external magnetic field has a large out-of-plane component, the vortex formation may cause problems in observing the angular dependence.
The qualitative behavior is expected to change when the out-of-plane magnetic field approaches the upper critical field ($H \sim H_{\mathrm c 2} \sim 1 \mathrm T$).
This is because the coherence length of the Cooper pair and the distance between the vortices can become comparable.
Indeed, it has been experimentally reported that the vortex formation suppresses the characteristic properties in the spin pumping into SCs \cite{jeonAbrikosovVortexNucleation2019}.
Therefore, the out-of-plane magnetic field should be as small as possible when FMR measurements are performed for $H \sim H_{\mathrm c 2}$.

Recent experiments have reported that $\mathrm{UTe}_2$ is a candidate material for spin-triplet $p$-wave SCs \cite{ranNearlyFerromagneticSpintriplet2019a}, which has attracted a great deal of attention. Various experiments, including spectroscopic measurements, are now in progress to investigate the pairing symmetry of $\mathrm{UTe}_2$, and indicated that the superconducting transition temperature is about 1K $\sim$ 30 GHz. 
Therefore, the resonance condition $\hbar \omega = 2 \Delta$ shown above is accessible to recent broadband FMR measurements.

In addition, experiments on spin pumping into $d$-wave SCs have recently been reported \cite{carreira2021spin} and a theoretical investigation of the enhancement of the Gilbert damping in a $d$-wave SC/FI bilayer system has recently been presented \cite{ominato2022ferromagnetic}.
Thus anisotropic superconducting spintronics can be expected to develop as a new research direction.

We should emphasize two important aspects of the FMR method presented here: the spectroscopic probe method for the $p$-wave SC thin films and the versatile spin injection method.
First, the FMR measurement procedure can provide a new spin-sensitive measurement method that will complement other measurement methods to enable a breakthrough in the discovery of spin-triplet SCs.
Second, the FMR method represents a promising way to generate spin-triplet currents in $p$-wave SC thin films.

{\it Conclusions.---}We have investigated the anisotropic superconducting spin transport at magnetic interfaces composed of a $p$-wave SC and an FI based on a microscopic model Hamiltonian.
The FMR signal in these $p$-wave SC/FI bilayer systems is modulated via spin transfer at the interface, which generates spin-triplet currents.
We have shown that the pairing symmetry of the SCs can be extracted from the FMR modulation characteristics.
Our approach provides a unique way to explore anisotropic superconducting spintronics, which will be useful for application to emerging device technologies.

{\it Note added.---} After the submission of this manuscript, we became aware of a closely related work, where a way to convert spin-triplet currents to magnon spin currents in SC/FI bilayer systems is discussed \cite{johnsenMagnonSpinCurrent2021a}.

We thank R. Ohshima, M. Shiraishi, H. Chudo, G. Okano, K. Yamanoi, and Y. Nozaki for helpful discussions.
This work was supported by the Priority Program of the Chinese Academy of Sciences under Grant No. XDB28000000, and by JSPS KAKENHI under Grants Nos. JP20K03835, JP20H04635, JP20H01863, JP21H04565, and JP21H01800.





\begin{widetext}

\section*{Supplemental Material}

\section{Model Hamiltonian}

In this section, we describe the derivation and details of the model Hamiltonian used in the main text.

\subsection{Ferromagnetic Heisenberg model}

\renewcommand{\theequation}{S.\arabic{equation}}
\setcounter{equation}{0}

The ferromagnetic Heisenberg model with the transverse AC magnetic field due to the microwave radiation is given by
\begin{align}
    H_{\mathrm{FI}}(t)
    =
    -J\sum_{\la i,j\ra}\bm{S}_i\cdot\bm{S}_j
    +\hbar\gamma H\sum_{j}S^Z_j
    -\hbar\gamma h_{\mathrm{ac}}\sum_{j}\left(S^X_j\cos\omega t-S^Y_j\sin\omega t\right),
\end{align}
where $J>0$ is the exchange coupling constant, $\la i,j\ra$ represents summation over all nearest-neighbor sites, $\bm{S}_j$ is the localized spin at site $j$ in the ferromagnetic insulator (FI), $\gamma(<0)$ is the gyromagnetic ratio, $H$ is a static magnetic field, $h_{\mathrm{ac}}$ is an amplitude of an transverse oscillating magnetic field due to the microwave radiation with a frequency $\omega$.
The rotated coordinates ($X,Y,Z$) are shown in Fig.\ \ref{fig_system}(a).

It is convenient to introduce the boson creation and annihilation operators in order to formulate the problem in terms of the quantum field theory.
In the current problem, we perturbatively treat the excitation of the FI.
In this case, the Holstein-Primakoff transformation is useful, where the localized spin can be described using boson creation and annihilation operators $b_j,b_j^\dagger$ in Hilbert space constrained to $2S+1$ dimensions.
The spin operators are written as
\begin{align}
    &S^+_j=S^X_j+iS^Y_j=\left(2S-b_j^\dagger b_j\right)^{1/2}b_j, \\
    &S^-_j=S^X_j-iS^Y_j=b_j^\dagger\left(2S-b_j^\dagger b_j\right)^{1/2}, \\
    &S^Z_j=S-b_j^\dagger b_j,
\end{align}
where we require $[b_i,b_j^\dagger]=\delta_{i,j},$ in order that the $S^+_j$, $S^-_j$, and $S_j^Z$ satisfy the commutation relation of angular momentum.
The deviation of $S^Z_j$ from its ground-state value $S$ is quantified by the boson particle number.

We consider low-energy excitation in the FI, where the deviation of $S^Z_j$ from the ground state is small $\la b_j^\dagger b_j\ra/S\ll1$.
The ladder operators $S_j^\pm$ are approximated as
\begin{align}
    &S_j^+\approx(2S)^{1/2}b_j, \\
    &S_j^-\approx(2S)^{1/2}b_j^\dagger,
\end{align}
which is called spin-wave approximation.
Here, we define the magnon operators
\begin{align}
    &b_{\bk}=\frac{1}{\sqrt{N}}\sum_je^{-i\bk\cdot\br_j}b_j, \\
    &b_{\bk}^\dagger=\frac{1}{\sqrt{N}}\sum_je^{i\bk\cdot\br_j}b_j^\dagger,
\end{align}
where $N$ is the number of sites and $\bk=(k_x,k_y,k_z)$.
The inverse transformation is then given by
\begin{align}
    &b_j=\frac{1}{\sqrt{N}}\sum_{\bk}e^{i\bk\cdot\br_j}b_{\bk}, \\
    &b_j^\dagger=\frac{1}{\sqrt{N}}\sum_{\bk}e^{-i\bk\cdot\br_j}b_{\bk}^\dagger.
\end{align}
The magnon operators satisfy $[b_{\bk},b_{\bk^\prime}^\dagger]=\delta_{\bk,\bk^\prime}$ and describe the quantized collective excitations.
Using the spin-wave approximation and the magnon operators, the Hamiltonian $H_{\mathrm{FI}}(t)$ is written as
\begin{align}
    H_{\mathrm{FI}}(t)
    \approx
    &\sum_{\bk}\hbar\omega_{\bk}b_{\bk}^\dagger b_{\bk}
    -h^+_{\mathrm{ac}}(t)b_{\bk=\bm{0}}^\dagger
    -h^-_{\mathrm{ac}}(t)b_{\bk=\bm{0}},
\end{align}
where $\hbar\ok=Dk^2-\hbar\gamma H$ with $D=2JSa^2$ and the lattice constant $a$, $h^{\pm}_{\mathrm{ac}}(t)=\hbar\gamma h_{\mathrm{ac}}\sqrt{SN/2}e^{\mp i\omega t}$, and constant terms are omitted.

\subsection{BCS Hamiltonian}

We derive a mean-field Hamiltonian, which describes a bulk superconductor (SC), and we diagonalize the mean-field Hamiltonian with the Bogoliubov transformation.
At the end of this section, the spin density operators of the SC are written in terms of the Bogoliubov quasiparticle creation and annihilation operators.

We start with the effective Hamiltonian in momentum space
\begin{align}
    H_{\mathrm{SC}}=\sum_{\bk,s}\xi_{\bk}
        c^\dagger_{\bk s}c_{\bk s}
        +
        \frac{1}{2}\sum_{\bk,\bkp,s_1,s_2,s_3,s_4}
        V_{s_1,s_2,s_3,s_4}(\bk,\bkp)
        c^\dagger_{-\bk s_1}c^\dagger_{\bk s_2}
        c_{\bkp s_3}c_{-\bkp s_4},
\end{align}
where $\xi_\bk$ is the band energy measured relative to the chemical potential, and $c_{\bk s}^\dagger$ and $c_{\bk s}$ are the creation and annihilation operators of electrons with the wave vector $\bk=(k_x,k_y)$ and the $z$ component of the spin $s=\uparrow,\downarrow$.
The matrix elements satisfy
\begin{align}
    V_{s_1,s_2,s_3,s_4}(\bk,\bkp)
    &=-V_{s_2,s_1,s_3,s_4}(-\bk,\bkp), \\
    V_{s_1,s_2,s_3,s_4}(\bk,\bkp)
    &=-V_{s_1,s_2,s_4,s_3}(\bk,-\bkp),
\end{align}
because of the anticommutation relation of fermions, and
\begin{align}   
    V_{s_1,s_2,s_3,s_4}(\bk,\bkp)    
    &=V^{\ast}_{s_4,s_3,s_2,s_1}(\bkp,\bk),
\end{align}
because of the Hermitianity of the Hamiltonian.
We consider a mean-field, which is called a pair potential
\begin{align}
    &\Delta_{\bk,ss^\prime}
    =-\sum_{\bkp,s_3,s_4}
        V_{s^\prime,s,s_3,s_4}(\bk,\bkp)
        \la c_{\bkp s_3}c_{-\bkp s_4}\ra,
        \label{eq_gap_equation}
\end{align}
and its conjugate
\begin{align}
    &\Delta^{\ast}_{-\bk,ss^\prime}
    =\sum_{\bkp,s_1,s_2}
        V_{s_1,s_2,s^\prime,s}(\bkp,\bk)
        \la c^\dagger_{-\bkp s_1}
        c^\dagger_{\bkp s_2}\ra.
\end{align}
Here, we consider a mean-field approximation where the interaction term is replaced as follows
\begin{align}
    c^\dagger_{-\bk s_1}c^\dagger_{\bk s_2}
    c_{\bkp s_3}c_{-\bkp s_4}
    \to
    c^\dagger_{-\bk s_1}c^\dagger_{\bk s_2}
    \la c_{\bkp s_3}c_{-\bkp s_4} \ra
    +
    \la c^\dagger_{-\bk s_1}c^\dagger_{\bk s_2} \ra
    c_{\bkp s_3}c_{-\bkp s_4}
    -
    \la c^\dagger_{-\bk s_1}c^\dagger_{\bk s_2} \ra
    \la c_{\bkp s_3}c_{-\bkp s_4} \ra,
\end{align}
so that the interaction term is rewritten as
\begin{align}
    \sum_{\bk,\bkp,s_1,s_2,s_3,s_4}
    V_{s_1,s_2,s_3,s_4}(\bk,\bkp)
    c^\dagger_{-\bk s_1}c^\dagger_{\bk s_2}
    c_{\bkp s_3}c_{-\bkp s_4}
    \to
    \sum_{\bk,s_1,s_2}
    \left[
        \Delta_{\bk,s_1s_2}
        c^\dagger_{\bk s_1}c^\dagger_{-\bk s_2}
        -
        \Delta^{\ast}_{-\bk,s_1s_2}
        c_{-\bk s_1}c_{\bk s_2}
    \right],
\end{align}
where an constant term is omitted.
Consequently, we derive a mean-field Hamiltonian
\begin{align}
    H_{\mathrm{SC}}
        =\sum_{\bk,s}\xi_{\bk}
            c^\dagger_{\bk s}c_{\bk s}
            +
            \frac{1}{2}\sum_{\bk,s_1,s_2}
            \big[
                \Delta_{\bk,s_1s_2}
                c^\dagger_{ \bk s_1}
                c^\dagger_{-\bk s_2}
                -
                \Delta^{\ast}_{-\bk,s_1s_2}
                c_{-\bk s_1}
                c_{ \bk s_2}
            \big].
\end{align}
Using a four-component notation
\begin{align}
    &\bm{c}_{\bk}^\dagger
    =
    (
        c_{\bk\uparrow}^\dagger,
        c_{\bk\downarrow}^\dagger,
        c_{-\bk\uparrow},
        c_{-\bk\downarrow}
    ), \\
    &\bm{c}_{\bk}
    =
    (
        c_{\bk\uparrow},
        c_{\bk\downarrow},
        c_{-\bk\uparrow}^\dagger,
        c_{-\bk\downarrow}^\dagger
    )^{\rm T},
\end{align}
the mean-field Hamiltonian is written as
\begin{align}
    H_{\mathrm{SC}}=
        \frac{1}{2}\sum_{\bk}
        \bm{c}_{\bk}^{\dagger}
        H_{\rm BdG}\bm{c}_{\bk}.
\end{align}
$H_{\rm BdG}$ is the $4\times 4$ matrix
\begin{align}
    H_{\rm BdG}=
        \begin{pmatrix}
             \xi_{\bk}\sigma^0   & \Delta_{\bk} \\
            -\Delta^{\ast}_{-\bk} & -\xi_{\bk}\sigma^0
        \end{pmatrix},
\end{align}
where $\sigma^0$ is the $2\times2$ unit matrix and $\Delta_{\bk}$ is the $2\times2$ matrix given as
\begin{align}
    \Delta_{\bk}=
    \begin{pmatrix}
        \Delta_{\bk,\uparrow\uparrow} & \Delta_{\bk,\uparrow\downarrow} \\
        \Delta_{\bk,\downarrow\uparrow} & \Delta_{\bk,\downarrow\downarrow}
    \end{pmatrix}.
\end{align}

In principle, the pair potential is obtained by solving the gap equation self-consistently for an explicit form of the matrix elements $V_{s_1,s_2,s_3,s_4}(\bk,\bkp)$.
In this work, we do not solve the gap equation, but instead assume an explicit form of the pair potential and perform calculations using a phenomenological gap function.
For the singlet pairing, the pair potential is given by
\begin{align}
    \Delta_{\bk}=\psi_{\bk}i\sigma^y,
\end{align}
with an even function $\psi_{\bk}=\psi_{-\bk}$.
For an $s$-wave SC, the pair potential is given by
\begin{align}
    \Delta_\bk=\Delta
    \begin{pmatrix}
        0  & 1 \\
        -1 & 0    
    \end{pmatrix}.
\end{align}
For the triplet pairing, the pair potential is given by
\begin{align}
    \Delta_\bk=[\bm{d}_{\bk}\cdot\bm{\sigma}]i \sigma^y,
\end{align}
with an odd vectorial function $\bm{d}_\bk=-\bm{d}_{-\bk}$.
For a chiral $p$-wave SC and a helical $p$-wave SC, $\bm{d}_{\bk}$ is given by
\begin{align}
    \bm{d}_{\bk}=
        \begin{cases}
            \Delta(0,0,e^{i\phi_\bk})            & :\mathrm{chiral} {~}p-{\rm wave} \\
            \Delta(-\sin\phi_\bk,\cos\phi_\bk,0) & :\mathrm{helical}{~}p-{\rm wave}
        \end{cases}
\end{align}
with $\phi_\bk=\arctan(k_y/k_x)$, so that the pair potential is given by
\begin{align}
    \Delta_\bk=
            \begin{cases}
                \Delta
                \begin{pmatrix}
                    0             & e^{i\phi_\bk} \\
                    e^{i\phi_\bk} & 0    
                \end{pmatrix} & :{\rm chiral} {~}p-{\rm wave} \\
                \Delta
                \begin{pmatrix}
                    ie^{-i\phi_\bk} & 0              \\
                    0               & ie^{i\phi_\bk}
                \end{pmatrix} & :{\rm helical}{~}p-{\rm wave}
            \end{cases}
\end{align}
The phenomenological gap function is given by
\begin{align}
    \Delta=1.76k_{\rm B}T_c\tanh\left(1.74\sqrt{T_c/T-1}\right).
\end{align}

The Bogoliubov transformation to diagonalize $H_{\mathrm{BdG}}$ is given by
\begin{align}
    &U_{\bk}=
        \begin{pmatrix}
            u_{\bk}         & v_{\bk}          \\
            v_{-\bk}^{\ast} & u_{-\bk}^{\ast}
        \end{pmatrix}, \\
    &U_{\bk}^\dagger=
        \begin{pmatrix}
            u_{\bk}          & -v_{\bk}          \\
            -v_{-\bk}^{\ast} & u_{-\bk}^{\ast}
        \end{pmatrix},
\end{align}
with the $2\times2$ matrices $u_\bk$ and $v_\bk$ given by
\begin{align}
    &u_{\bk}=
        \sqrt{
            \frac{1}{2}
            \left(
                1+\frac{\xi_\bk}{E_\bk}
            \right)
            }\sigma^0, \\
    &v_{\bk}=
        -\sqrt{
            \frac{1}{2}
            \left(
                1-\frac{\xi_\bk}{E_\bk}
            \right)
            }
        \frac{\Delta_{\bk}}{\Delta},
\end{align}
where $E_\bk$ is the eigenenergy
\begin{align}
    E_{\bk}=\sqrt{\xi_{\bk}^2+\Delta^2}.
\end{align}
Using the Bogoliubov transformation $U_\bk$, the $4\times4$ matrix $H_{\rm BdG}$ is diagonalized as
\begin{align}
    U_{\bk}^\dagger H_{\rm BdG} U_{\bk}=
        \begin{pmatrix}
            E_{\bk} & 0       & 0        & 0         \\
            0       & E_{\bk} & 0        & 0         \\
            0       & 0       & -E_{\bk} & 0         \\
            0       & 0       & 0        & -E_{\bk}
        \end{pmatrix}.
\end{align}
The excitation of $H_{\mathrm{SC}}$ is described by the creation and annihilation operators of the Bogoliubov quasiparticles $\bm{\gamma}_\bk^{(\dagger)}$
\begin{align}
    \bm{\gamma}_{\bk}^\dagger&=
        (
            \gamma_{\bk\uparrow}^\dagger,
            \gamma_{\bk\downarrow}^\dagger,
            \gamma_{-\bk\uparrow},
            \gamma_{-\bk\downarrow}
        ), \\
    \bm{\gamma}_{\bk}&=
        (
            \gamma_{ \bk\uparrow  },        
            \gamma_{ \bk\downarrow},         
            \gamma_{-\bk\uparrow  }^{\dagger},
            \gamma_{-\bk\downarrow}^{\dagger}
        )^{\mathrm{T}},
\end{align}
where they are obtained by the Bogoliubov transformation
\begin{align}
        &\bm{\gamma}_{\bk}=U_{\bk}^\dagger\bm{c}_{\bk}, \\
        &\bm{\gamma}_{\bk}^\dagger=\bm{c}_{\bk}^\dagger U_{\bk}.
\end{align}

The spin density operators $\sigma^a(\br)$ ($a=x,y,z$) is defined as
\begin{align}
    \sigma^a(\br):=
           \frac{1}{A}
           \sum_{\bk,\bkp,s,s^\prime}
            e^{-i(\bk-\bkp)\cdot\br}
            \sigma^a_{ss^\prime}
            c^\dagger_{\bk s}
            c_{\bkp s^\prime},
\end{align}
where $A$ is the area of the system.
$\sigma^a(\br)$ $(a=x,y,z)$ is expanded in Fourier series
\begin{align}
    &\sigma^a(\br)=\frac{1}{A}\sum_\bq e^{i\bq\cdot\br}\sigma^a_\bq,
\end{align}
and the Fourier coefficient is given by
\begin{align}
    \sigma^a_\bq
        =\int d\br e^{-i\bq\cdot\br}\sigma^a(\br)
        =\sum_{\bk,s,s^{\prime}}
         \sigma^a_{ss^\prime}
         c^{\dagger}_{\bk s}
         c_{\bk+\bq s^\prime}.
\end{align}
Using the Bogoliubov transformation $U_{\bk}$, the above expression is rewritten as
    \begin{align}
        \sigma^a_{\bq}=
            \sum_{\bk,s,s^\prime}
            \Biggl[
            &\left(
                s_{\bk,\bk+\bq}^{a(1)}
            \right)_{s,s^\prime}
            \gamma_{\bk     s       }^\dagger
            \gamma_{\bk+\bq s^\prime}
            +
            \left(
                s_{\bk,\bk+\bq}^{a(2)}
            \right)_{s,s^\prime}
            \gamma_{-\bk     s       }
            \gamma_{-\bk-\bq s^\prime}^\dagger
            +
            \left(
                s_{\bk,\bk+\bq}^{a(3)}
            \right)_{s,s^\prime}
            \gamma_{ \bk     s       }^\dagger
            \gamma_{-\bk-\bq s^\prime}^\dagger
            +
            \left(
                s_{\bk,\bk+\bq}^{a(4)}
            \right)_{s,s^\prime}
            \gamma_{-\bk     s       }
            \gamma_{ \bk+\bq s^\prime}
            \Biggr],
        \label{eq_spin_q}
    \end{align}
with the $2\times2$ matrices $s_{\bk,\bk+\bq}^{a(i)}$ given by
\begin{align}
    s_{\bk,\bk+\bq}^{a(1)}=
        u_\bk^\dagger
        \sigma^a
        u_{\bk+\bq}, \\
    s_{\bk,\bk+\bq}^{a(2)}=
        v_\bk^\dagger
        \sigma^a
        v_{\bk+\bq}, \\
    s_{\bk,\bk+\bq}^{a(3)}=
        u_\bk^\dagger
        \sigma^a
        v_{\bk+\bq}, \\
    s_{\bk,\bk+\bq}^{a(4)}=
        v_\bk^\dagger
        \sigma^a
        u_{\bk+\bq}.
\end{align}
The first and second terms describe the intraband transition from particle-to-particle and from hole-to-hole, respectively.
The third and fourth terms describe the interband transition from hole-to-particle and from particle-to-hole, respectively.

\subsection{Proximity exchange coupling at interface}

We start with a model for the proximity exchange coupling given by
\begin{align}
    \hex=
        \int d\br
        \sum_{j}
        J(\br,\br_j)
        \bm{\sigma}(\br)\cdot\bm{S}_j.
\end{align}
We rewrite the above expression in the real space into the expression in the wave space.
The proximity exchange coupling is rewritten as
\begin{align}
    \hex=
        \int d\br
        \sum_{j}
        J(\br,\br_j)
        \frac{1}{A\sqrt{N}}
        \sum_{\bq,\bk}
        e^{i(\bq\cdot\br+\bk\cdot\br_j)}
        \left(
            \sigma_\bq^+S_\bk^-+\sigma_\bq^-S_\bk^+
        \right)
        +\int d\br
        \sum_{j}
        J(\br,\br_j)
        \sigma^Z(\br)S^Z_j,
\end{align}
where the Fourier series are given by
\begin{align}
        &\bm{\sigma}(\br)=
            \frac{1}{A}
            \sum_\bq
            e^{i\bq\cdot\br}
            \bm{\sigma}_\bq,  \\
        &\bm{S}_j=
            \frac{1}{\sqrt{N}}
            \sum_\bk
            e^{i\bk\cdot\br_j}
            \bm{S}_\bk,
\end{align}
with the area of the SC, $A$, and the number of sites in the FI, $N$, and the ladder operators are given by
\begin{align}
    &\sigma^\pm=\frac{1}{2}(\sigma^X\pm i\sigma^Y), \\
    &S^\pm=S^X\pm iS^Y.
\end{align}
The matrix element is given by
\begin{align}
    J_{\bq,\bk}=
        \frac{1}{A\sqrt{N}}
        \int d\br
        \sum_{j}
        J(\br,\br_j)
        e^{i(\bq\cdot\br+\bk\cdot\br_j)}.
\end{align}
Consequently, the exchange coupling which we use in the main text is derived as
\begin{align}
    \hex=
    \sum_{\bq,\bk}
    \left(
        J_{\bq,\bk}\sigma_\bq^+S_\bk^-
        +
        J_{\bq,\bk}^\ast\sigma_{-\bq}^-S_{-\bk}^+
    \right),
\end{align}
where we use a relation $J_{-\bq,-\bk}=J_{\bq,\bk}^\ast$, and we omit the last term
\begin{align}
    \int d\br\sum_jJ(\br,\br_j)\sigma^Z(\br)S_j^Z,
\end{align}
in order to focus on the spin transfer at the interface.
For the uniform magnon mode $|\bk|=0$, the matrix element is given by
\begin{align}
    J_{\bq,\bk=\bm{0}}=
        \frac{1}{A\sqrt{N}}
        \int d\br
        \sum_{j}
        J(\br,\br_j)
        e^{i\bq\cdot\br}.
        \label{eq_mat_ele}
\end{align}

\section{Time dependent quantum average}

In this section, we show that the ferromagnetic resonance (FMR) frequency and linewidth are read from the magnon Green's function.
We consider the Hamiltonian $H(t)$ composed of the unperturbed Hamiltonian $H_0$ and the perturbation $V(t)$
\begin{align}
    H(t)=H_0+V(t).
\end{align}
The time-dependent quantum average of a physical quantity $O$ is calculated as
\begin{align}
\la O(t)\ra
    &=\la
        \S^\dagger(t,-\infty)
        \tilde{O}(t)
        \S(t,-\infty)
    \ra,
\end{align}
where $\tilde{O}(t)$ is the interaction picture and the S matrix $\S(t,t_0)$ is given by
\begin{align}
    \S(t,t_0)=T\exp\left(\int^t_{t_0}dt^\prime\frac{\tilde{V}(t^\prime)}{i\hbar}\right).    
\end{align}
The time-dependent quantum average $\la O(t) \ra$ is written as
\begin{align}
    \la O(t) \ra = \la O \ra_{\rm eq} 
                    + \delta\la O(t) \ra,
\end{align}
where $\la O \ra_{\rm eq}={\rm Tr}\left(\rho_{\rm eq} O\right)$ is the equilibrium value and $\delta\la O(t) \ra$ is deviation from the equilibrium.
When the perturbation is written as $V(t)=-AF(t)$, the first order perturbation calculation gives
\begin{align}
    \d\la O(t)\ra
        &=-\int^t_{-\infty}dt^\p\frac{1}{i\hbar}
        \la[\tilde{O}(t),\tilde{A}(t^\p)]\ra F(t^\p) \notag \\
        &=-\int^\infty_{-\infty}dt^\p G^R(t^\p)F(t-t^\p),
\end{align}
where we define the retarded Green's function
\begin{align}
    G^R(t)=
        \frac{1}{i\hbar}
        \theta(t)
        \la[\tilde{O}(t),\tilde{A}(0)]\ra.
\end{align}
When the external force is written as $F(t)=Fe^{-i(\omega+i0) t}$, $\d\la O(t)\ra$ is written as
\begin{align}
    \d\la O(t)\ra
        &=-Fe^{-i(\omega+i0) t}\int^\infty_{-\infty}dt^\p e^{i(\omega+i0) t^\p}G^R(t^\p) \notag \\
        &=-Fe^{-i\omega t}G^R(\omega).
\end{align}
Using the above formula, the dynamics of $\d\la S^+_{\bk=\bm{0}}(t)\ra$ is written as
\begin{align}
    \d\la S^+_{\bk=\bm{0}}(t)\ra=
        -\frac{\hbar\gamma\hac\sqrt{N}}{2} e^{-i\omega t}
        G^R_{\bk=\bm{0}}(\omega),
        \label{eq_spin_dynamics}
\end{align}
where $G^R_\bk(\omega)$ is the Fourier transform of the retarded component of the magnon Green's function $G^R_\bk(t)$. They are defined as
\begin{align}
    &G^R_\bk(t):=
        \frac{1}{i\hbar}
        \theta(t)
        \la[
            S^+_{\bk}(t),
            S^-_{-\bk}(0)
        ]\ra, \\
    &G^R_\bk(\omega):=
        \int^\infty_{-\infty}dt^\p
        e^{i(\omega+i0) t^\prime}
        G^R_\bk(t^\prime).
\end{align}
From Eq.\ (\ref{eq_spin_dynamics}), one can see that the FMR frequency and linewidth are read from $G^R_\bk(\omega)$.

\section{Magnon Green's function}

In this section, we perform perturbative calculation for the magnon Green's function. We treat the proximity exchange coupling as a perturbation. The Hamiltonian is written as
\begin{align}
    H=H_0+V,
\end{align}
where $H_0$ is the unperturbed Hamiltonian
\begin{align}
    H_0=
        \sum_\bk
        \hbar\omega_\bk
        b_\bk^\dagger
        b_\bk
        +
        \sum_{\bk,s}
        E_{\bk}
        \gamma_{\bk s}^\dagger
        \gamma_{\bk s},
\end{align}
and $V$ is the perturbation
\begin{align}
    V&=\sum_{\bq,\bk}
        \left(
            J_{\bq,\bk}
            \sigma_\bq^+
            S_\bk^-
            +{\rm h.c.}
        \right).
\end{align}

\begin{figure}
    \begin{center}
    \includegraphics[width=1\hsize]
    {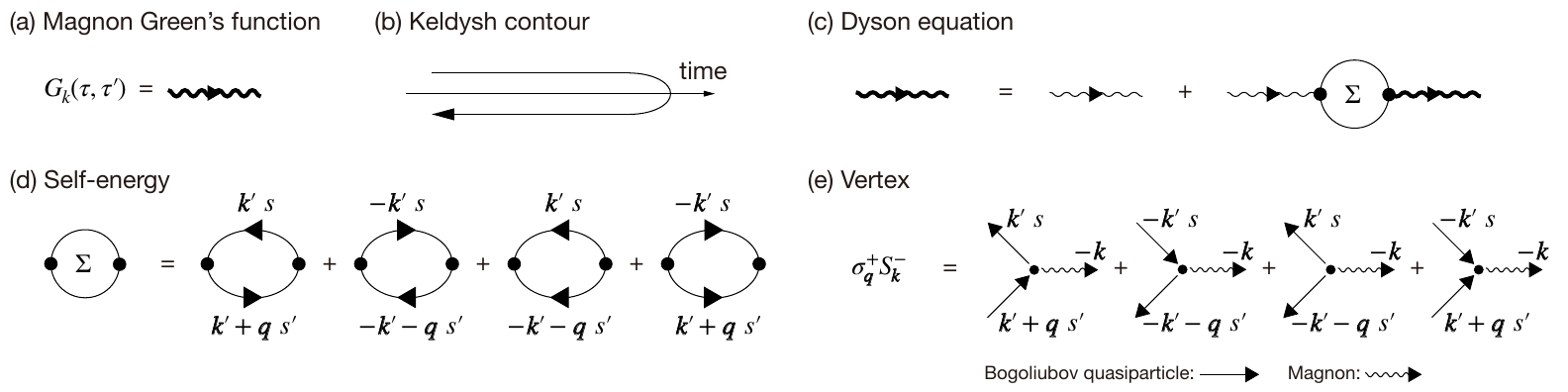}
    \end{center}
    \caption{
    (a) The Feynman diagram for the magnon Green's function.
    (b) Keldysh contour to perform perturbative calculations.
    (c) The Feynman diagram for the Dyson equation.
    (d) The self-energy within the second-order perturbation is given by the dynamic spin susceptibility of the SCs.
    (e) The Feynman diagrams for the vertex $\sigma^+_{\bq}S^-_{\bk}$, which represent scattering of a Bogoliubov quasiparticle with magnon emission. The solid and wavy lines represent a Bogoliubov quasiparticle and a magnon, respectively.}
    \label{fig_diagram}
\end{figure}

We define the magnon Green's function
\begin{align}
    G_{\bk}(\tau,\tau^\prime):=
        \frac{1}{i\hbar}
        \la
            T_C
            S_{ \bk}^+(\tau       )
            S_{-\bk}^-(\tau^\prime)
        \ra,
\end{align}
where $T_C$ is the time-ordering operator on the Keldysh contour (see Figs.\ \ref{fig_diagram}(a) and (b)).
To perform the perturbative calculation, we introduce interaction picture.
The perturbation is written as
\begin{align}
    \tilde{V}(t)
        &=\sum_{\bq,\bk}
            \left(
                J_{\bq,\bk}
                \tilde{\sigma}_{\bq}^+(t)
                \tilde{S}^-_\bk(t)
                +{\rm h.c.}
            \right).
\end{align}
The magnon Green's function is given by
\begin{align}
    G_{\bk}(\tau,\tau^\prime)
        &=\frac{1}{i\hbar}
            \la
                T_C\S_C
                \tilde{S}^+_{ \bk}(\tau)
                \tilde{S}^-_{-\bk}(\tau^\prime)
            \ra_{\mathrm{conn}},
\end{align}
where $\la\cdots\ra_{\mathrm{conn}}$ means the connected diagrams and the S matrix is given by
\begin{align}
    \S_C=T_C\exp\left(\int_Cd\tau\frac{\tilde{V}(\tau)}{i\hbar}\right).
\end{align}
The above expressions lead to the Dyson equation (see Fig.\ \ref{fig_diagram}(c))
\begin{align}
    G_\bk(\tau,\tau^\prime)
        =G^{(0)}_\bk(\tau,\tau^\prime)
        +\int_Cd\tau_1\int_Cd\tau_2
            G^{(0)}_\bk(\tau,\tau_1)
            \Sigma_\bk(\tau_1,\tau_2)
            G_\bk(\tau_2,\tau^\prime),
\end{align}
where $G^{(0)}_\bk(\tau,\tau^\prime)$ is the unperturbed magnon Green's function 
\begin{align}
    G^{(0)}_\bk(\tau,\tau^\prime)
        =\frac{1}{i\hbar}
        \la
            T_C
            \tilde{S}_{ \bk}^+(\tau       )
            \tilde{S}_{-\bk}^-(\tau^\prime)
        \ra,
\end{align}
and $\Sigma_\bk(\tau_1,\tau_2)$ is the self-energy. Within the second-order perturbation, the self-energy is given by (see Fig.\ \ref{fig_diagram}(d))
\begin{align}
    \Sigma_\bk(\tau,\tau^\prime)=\frac{1}{i\hbar}
        \sum_{\bq}
        |J_{\bq,\bk}|^2
        \la
            T_C
            \tilde{\sigma}_{ \bq}^+(\tau)
            \tilde{\sigma}_{-\bq}^-(\tau^\prime)
        \ra.
\end{align}
The Feynman diagram for the vertex is shown in Fig.\ \ref{fig_diagram}(e).
Substituting the ladder operators expressed in terms of $\gamma_{\bk s}^{(\dagger)}$, the self-energy is written as
    \begin{align}
        \Sigma_\bk(\tau,\tau^\prime)=-i\hbar
            \sum_{\bq}
            \abs{J_{\bq,\bk}}^2
            \sum_{\bkp,s,s^\prime}
            \Bigg[
            &\qty(
                \abs{(s^{+(1)}_{\bkp,\bkp+\bq})_{s,s^\prime}}^2
                -
                (s^{+(1)}_{\bkp,\bkp+\bq})_{s,s^\prime}
                (s^{-(2)}_{-\bkp-\bq,-\bkp})_{s^\prime,s}^\ast
            )
            g_{\bkp,s}(\tau^\prime,\tau)
            g_{\bkp+\bq,s^\prime}(\tau,\tau^\prime) \notag \\
            +
            &\qty(
                \abs{(s^{+(2)}_{\bkp,\bkp+\bq})_{s,s^\prime}}^2
                -
                (s^{+(2)}_{\bkp,\bkp+\bq})_{s,s^\prime}
                (s^{-(1)}_{-\bkp-\bq,-\bkp})_{s^\prime,s}^\ast
            )
            g_{-\bkp,s}(\tau,\tau^\prime)
            g_{-\bkp-\bq,s^\prime}(\tau^\prime,\tau) \notag \\
            -
            &\qty(    
                \abs{(s^{+(3)}_{\bkp,\bkp+\bq})_{s,s^\prime}}^2
                -
                (s^{+(3)}_{\bkp,\bkp+\bq})_{s,s^\prime}
                (s^{-(3)}_{-\bkp-\bq,-\bkp})_{s^\prime,s}^\ast
            )
            g_{ \bkp,s}(\tau^\prime,\tau)
            g_{-\bkp-\bq,s^\prime}(\tau^\prime,\tau) \notag \\
            -
            &\qty(
                \abs{(s^{+(4)}_{\bkp,\bkp+\bq})_{s,s^\prime}}^2
                -
                (s^{+(4)}_{\bkp,\bkp+\bq})_{s,s^\prime}
                (s^{-(4)}_{-\bkp-\bq,-\bkp})_{s^\prime,s}^\ast
            )
            g_{-\bkp,s}(\tau,\tau^\prime)
            g_{ \bkp+\bq,s^\prime}(\tau,\tau^\prime)
            \Bigg],
            \label{eq_self_energy_gg}
    \end{align}
where the quasiparticle Green's function is defined as
\begin{align}
    g_{\bk,s}(\tau,\tau^\prime)
    :=
    \frac{1}{i\hbar}
    \la
    T_C
    \tilde{\gamma}_{\bk s}(\tau)
    \tilde{\gamma}_{\bk s}^\dagger(\tau^\prime)
    \ra.
\end{align}
The first and second terms give the intraband contribution, and the third and fourth terms give the interband contribution.
Evaluating the Dyson equation, the retarded component of the magnon Green's function is given by
\begin{align}
    &G_\bk^R(\omega)
        =\frac
            {1}
            {
            \left[
                G^{(0)R}_\bk(\omega)
            \right]^{-1}
            -\Sigma^R_\bk(\omega)
            },
\end{align}
where the unperturbed Green's function is written as
\begin{align}
    &G^{(0)R}_\bk(\omega)
        =\frac
        {2S/\hbar}
        {
            \omega
            -\omega_\bk
            +i\alpha\omega
        }. \label{unperturbedGreen}
\end{align}
Here, we introduce the phenomenological dimensionless damping parameter $\alpha$.
Using Eq.(\ref{unperturbedGreen}), the retarded Green's function is written as
\begin{align}
    G_\bk^R(\omega)
        &=\frac
            {2S/\hbar}
            {
                \omega
                -\omega_\bk
                +i\alpha\omega
                -(2S/\hbar)\Sigma^R_\bk(\omega)
            }.
\end{align}
From the above expression, the frequency shift at a fixed $\omega$ is given by
\begin{align}
    \delta H=\frac{2S}{\gamma\hbar}{\rm Re}\Sigma^R_{\bk}(\omega),
\end{align}
and the enhanced Gilbert damping is given by
\begin{align}
    \delta\alpha=-\frac{2S}{\hbar\omega}\mathrm{Im}\Sigma^R_{\bk}(\omega).
\end{align}
The Fourier transform of the self-energy is given as
\begin{align}
    \Sigma^R_{\bk}(\omega)
    =
    \int dte^{i(\omega+i0) t}\Sigma^R_{\bk}(t)
    =
    -\sum_{\bq}
        |J_{\bq,\bk}|^2
        \chi_{\bq}^{R}(\omega),
    \label{eq_self_energy}
\end{align}
where the dynamic spin susceptibility of the SC is defined as
\begin{align}
    &\chi_{\bq}^{R}(\omega):=
        \int dte^{i(\omega+i0)t}
        \frac{i}{\hbar}
        \theta(t)
        \la[
            \tilde{\sigma}_{ \bq}^+(t),
            \tilde{\sigma}_{-\bq}^-(0)
        ]\ra.
\end{align}
Evaluating the self-energy Eq.\ (\ref{eq_self_energy}), one can obtain the information of the FMR modulation, $\delta H$ and $\delta\alpha$.
Using the system's symmetry, the dynamic spin susceptibility $\chi_\bq^{R}(\omega)$ can be written as
\begin{align}
    \chi_\bq^{R}(\omega)
        &=\cos^2\theta\chi_\bq^{xx}(\omega)
            +
            \chi_\bq^{yy}(\omega)
            +
            \sin^2\theta\chi_\bq^{zz}(\omega),
\end{align}
which means that both $\delta H$ and $\delta\alpha$ show a dependence on $\theta$ when the dynamic spin susceptibility is anisotropic.

\section{Spin current at the interface}

In this section, we derive the general expression of spin current at the interface.
We treat the tunneling Hamiltonian as a perturbation and the other terms as the unperturbed Hamiltonian
\begin{align}
    &H(t)=H_0(t)+H_{\mathrm{ex}}, \\
    &H_0(t)=\hfi(t)+H_{\mathrm{SC}}.
\end{align}
The operator of spin current flowing from the SC to the FI at the interface is defined by
\begin{align}
    \IS:=-\frac{\hbar}{2}\dot{\sigma}^Z_{\mathrm{tot}}=
        -\frac{\hbar}{2}
        \frac{1}{i\hbar}[\sigma^Z_{\mathrm{tot}},H_{\mathrm{ex}}]
        =\frac{i}{2}[\sigma^Z_{\mathrm{tot}},H_{\mathrm{ex}}],
\end{align}
where $\sigma^Z_{\mathrm{tot}}$ is given by
\begin{align}
    \sigma^Z_{\mathrm{tot}}=\int d\br\sigma^Z(\br).
\end{align}
Calculating the commutation relation, we obtain the following expression
\begin{align}
	\IS=
		i\sum_{\bq,\bk}
		\left(
			J_{\bq,\bk}
			\sigma_\bq^+
			S_\bk^-
			-
			\rm{h.c.}
		\right).
\end{align}
The time-dependent quantum average of $\IS$ is written as
\begin{align}
\la\IS(t)\ra
&={\rm Re}
	\left[
        2i
        \sum_{\bq,\bk}
		J_{\bq,\bk}
		\la
			\sigma_\bq^+(t)
			S_\bk^-(t)
		\ra
    \right],
\end{align}
where $\la\cdots\ra={\rm Tr}[\rho_0\cdots]$ denotes the statistical average with an initial density matrix $\rho_0$.
In order to develop the perturbation expansion, we introduce the interaction picture
\begin{align}
	\la
		\IS(\tau_1,\tau_2)
	\ra
	&=
	{\rm Re}
	\left[
        2i
        \sum_{\bq,\bk}
		J_{\bq,\bk}
		\la
			T_C
			\S_C
			\ts_\bq^+(\tau_1)
			\tS_\bk^-(\tau_2)
		\ra
	\right].
\end{align}
$\S_C$ and $\tilde{O}(t)$ are given by
    \begin{align}
        \S_C=
        T_C\exp
        \left(
            \int_C
            d\tau
            \frac
            {\tilde{H}_{\mathrm{ex}}(\tau)}
            {i\hbar}
        \right),
    \end{align}
and
    \begin{align}
        \tilde{O}(t)=
        \U0^\dagger(t,t_0)O\U0(t,t_0),
    \end{align}
where
    \begin{align}
        \U0(t,t_0)=
        T\exp
        \left(
            \int_{t_0}^t
            dt^\p
            \frac{H_0(t^\p)}{i\hbar}
        \right).
    \end{align}
Expanding $\S_C$ as
    \begin{align}
        \S_C\approx
            1+
            \int_Cd\tau
            T_C
            \frac
            {\tilde{H}_{\mathrm{ex}}(\tau)}
            {i\hbar},
    \end{align}
the spin current is given by
    \begin{align}
        \la\IS(\tau_1,\tau_2)\ra=
        \sum_{\bq,\bk}
        |J_{\bq,\bk}|^2
        {\rm Re}
        	\Bigg[
                \frac{2}{\hbar}
        		\int_Cd\tau
        		\la 
        			T_C
        			\ts_\bq^+(\tau_1)
        			\ts_{-\bq}^-(\tau)
        		\ra
        		\la 
        			T_C
        			\tS_{-\bk}^+(\tau)
        			\tS_\bk^-(\tau_2)
        		\ra
            \Bigg].
    \end{align}
Using the contour ordered Green's functions
    \begin{align}
        &\chi_{\bq}(\tau_1,\tau)=
        -\frac{1}{i\hbar}
        \la T_C
            \ts_{\bq}^+(\tau_1)
            \ts_{-\bq}^-(\tau)
        \ra, \\
        &G_{\bk}(\tau,\tau_2)=
        \frac{1}{i\hbar}
        \la T_C
            \tS_{ \bk}^+(\tau)
            \tS_{-\bk}^-(\tau_2)
        \ra,
    \end{align}
the above equation is rewritten as
    \begin{align}
        \la\IS(\tau_1,\tau_2)\ra
        =\sum_{\bq,\bk}|J_{\bq,\bk}|^2
        {\rm Re}
        	\Bigg[
                2\hbar
        		    \int_Cd\tau
        			\chi_{\bq}(\tau_1,\tau)
        			G_{-\bk}(\tau,\tau_2)
                \Bigg].
    \end{align}
We put $\tau_2$ on the forward contour and $\tau_1$ on the backward contour to describe spin transfer at the interface in appropriate time order.
Assuming a steady state, the spin current is written as
    \begin{align}
        \la\IS\ra=
        2\hbar
         \sum_{\bq,\bk}
         |J_{\bq,\bk}|^2
        {\rm Re}
    	    \Bigg[
                \int_{-\infty}^\infty
                \frac{d\omega^\prime}{2\pi}
	    	\Big(
	    		 \c_{ \bq}^R(\omega^\prime)
	    		  G_{-\bk}^<(\omega^\prime)
	    		+\c_{ \bq}^<(\omega^\prime)
	    		  G_{-\bk}^A(\omega^\prime)
	    	\Big)
    	\Bigg].
    \end{align}
We introduce the distribution functions as
    \begin{align}
        &\chi^<_\bq(\omega)=
        f^{\mathrm{SC}}_\bq(\omega)
        	\left[
        		2i{\mathrm{Im}}\chi^R_\bq(\omega)
        	\right], \\
        &G^<_\bk(\omega)=
        f^{\rm FI}_\bk(\omega)
        \left[
            2i{\mathrm{Im}}G^R_\bk(\omega)
        \right].
    \end{align}
The formula of the spin current at the interface is derived as
\begin{align}
	\la\IS\ra
	=
		4\hbar
		\sum_{\bq,\bk}
		|J_{\bq,\bk}|^2
		\int_{-\infty}^\infty
		\frac{d\omega^\prime}{2\pi}
		{\mathrm{Im}}\chi^R_{\bq}(\omega^\prime)
		\left[-{\mathrm{Im}}G^R_{-\bk}(\omega^\prime)\right]
		\left[
			f^{\rm FI}_{-\bk}(\omega^\prime)
			-
			f^{\rm SC}_{\bq}(\omega^\prime)
		\right].
\end{align}
When both the SC and the FI are in equilibrium, the difference of the distribution functions is zero (i.e. $f^{\mathrm{FI}}_{-\bk}(\omega^\prime)-f^{\mathrm{SC}}_{\bq}(\omega^\prime)=0$), so that no spin current is generated.
Under the microwave irradiation, the distribution function of the FI deviates from equilibrium, which generates the interface spin current.
Performing a second-order perturbation calculation, the deviation of the distribution function of the FI,  $\delta f^{\mathrm{FI}}_{-\bk}(\omega^\prime)$, is given by
\begin{align}
    \delta f^{\mathrm{FI}}_{-\bk}(\omega^\prime)
    =
    \frac{2\pi NS(\gamma h_{\mathrm{ac}}/2)^2}{\alpha \omega^\prime}
    \delta_{\bk,\bm{0}}\delta(\omega^\prime-\omega).
\end{align}
Consequently, the interface spin current is written as
\begin{align}
	\la\IS\ra^{\mathrm{SP}}=
		4\hbar
		\sum_{\bq,\bk}
		|J_{\bq,\bk}|^2
		\int_{-\infty}^\infty
		\frac{d\omega^\prime}{2\pi}
		{\mathrm{Im}}\chi^R_{\bq}(\omega^\prime)
		\left[-{\mathrm{Im}}G^R_{-\bk}(\omega^\prime)\right]
		\delta f^{\mathrm{FI}}_{-\bk}(\omega^\prime).
\end{align}
Finally, one can show that the spin current is proportional to the enhanced Gilbert damping
\begin{align}
    \la\IS\ra^{\rm SP}
    &=
    4\hbar\frac{NS(\gamma h_{\mathrm{ac}}/2)^2}{\alpha\omega}
    \left[-{\mathrm{Im}}G^R_{\bk=\bm{0}}(\omega)\right]
    \sum_{\bq}|J_{\bq,\bk=\bm{0}}|^2
    \mathrm{Im}\chi^{R}_{\bq}(\omega), \notag \\
    &=
    \frac{N(\hbar\gamma h_{\mathrm{ac}})^2}{2\alpha}
    \left[-{\mathrm{Im}}G^R_{\bk=\bm{0}}(\omega)\right]
    \delta\alpha.
\end{align}

\section{Model for interface configurations}

In order to calculate Eq.\ (\ref{eq_self_energy}), one needs to set up an explicit expression for $|J_{\bq,\bk=\bm{0}}|^2$.
We consider an interface with uncorrelated roughness.
To model this interface, we assume that $J(\br,\br_j)$ satisfies
\begin{align}
    \Big\la
        \sum_jJ(\br,\br_j)
    \Big\ra_{\mathrm{ave}}
    &=
    J_1, \label{eq_conf_ave1} \\
    \Big\la
        \sum_{j,j^\prime}J(\br,\br_j)J(\br^\prime,\br_{j^\prime})
    \Big\ra_{\mathrm{ave}}
    &=
    J_1^2+J_2^2l^2\delta(\br-\br^\prime), \label{eq_conf_ave2}
\end{align}
where $\la\cdots\ra_{\mathrm{ave}}$ means interface configuration average.
The spatially averaged $J(\br,\br_j)$ is given by a constant $J_1$ as shown in Eq.\ (\ref{eq_conf_ave1}).
Equation (\ref{eq_conf_ave2}) means that the interface roughness is uncorrelated and $J_2^2l^2$ is a variance.
$J_1$ and $J_2$ are coupling constants with dimension of energy, and are independent of the system size.
$l$ is introduced because the Hamiltonian of the SCs is treated as a continuum model.
Performing the interface configuration average, and using Eq.\ (\ref{eq_conf_ave1}) and (\ref{eq_conf_ave2}), one can obtain the expression for $|J_{\bq,\bk=\bm{0}}|^2$ in the main text.

\section{Dynamic spin susceptibility of SC}

Evaluating the retarded component of the self-energy Eq.\ (\ref{eq_self_energy_gg}), the dynamic spin susceptibility of the SC is given by
    \begin{align}
        \chi^{R}_{\bq}(\omega)=
            -\int^{\infty}_{-\infty}dE
            f(E)\sum_{\lambda,\bk}
            \Biggl\{
        &M^{\lambda,\lambda(a)}_{\bk,\bk+\bq}
            \left[
                -\frac{1}{\pi}{\rm Im}
                g^{R}_{\lambda,\bk}(E)
                g^{R}_{\lambda,\bk+\bq}(E+\hbar\omega)
                -\frac{1}{\pi}{\rm Im}
                g^{R}_{\lambda,\bk+\bq}(E)
                g^{A}_{\lambda,\bk}(E-\hbar\omega)
            \right] \notag \\
        +&M^{\lambda,-\lambda(a)}_{\bk,\bk+\bq}
            \left[
                -\frac{1}{\pi}{\rm Im}
                g^{R}_{ \lambda,\bk}(E)
                g^{R}_{-\lambda,\bk+\bq}(E+\hbar\omega)
                -\frac{1}{\pi}{\rm Im}
                g^{R}_{-\lambda,\bk+\bq}(E)
                g^{A}_{ \lambda,\bk}(E-\hbar\omega)
            \right]
        \Biggr\}, \label{eq_dynamic_spin_susceptibility_green_fn}
    \end{align}
where $M^{\lambda,\lambda^\prime(a)}_{\bk,\bk+\bq}$ with $a=s,c,$ and $h$ are given by
    \begin{align}
        &M^{\lambda,\lambda^\prime(s)}_{\bk,\bk+\bq}
            =
            \frac
            {(\xi_\bk+\lambda E_\bk)(\xi_{\bk+\bq}+\lambda^\prime E_{\bk+\bq})}
            {4\lambda E_{\bk}\lambda^\prime E_{\bk+\bq}}
            +
            \frac{\Delta^2}{4\lambda E_{\bk}\lambda^\prime E_{\bk+\bq}}, \\
        &M^{\lambda,\lambda^\prime(c)}_{\bk,\bk+\bq}
            =
            \frac
            {(\xi_\bk+\lambda E_\bk)(\xi_{\bk+\bq}+\lambda^\prime E_{\bk+\bq})}
            {4\lambda E_{\bk}\lambda^\prime E_{\bk+\bq}}
            -
            \frac
            {\Delta^2e^{-i(\phi_{\bk}-\phi_{\bk+\bq})}}
            {4\lambda E_{\bk}\lambda^\prime E_{\bk+\bq}}
            \cos^2\theta, \\
        &M^{\lambda,\lambda^\prime(h)}_{\bk,\bk+\bq}
            =
            \frac
            {(\xi_\bk+\lambda E_\bk)(\xi_{\bk+\bq}+\lambda^\prime E_{\bk+\bq})}
            {4\lambda E_{\bk}\lambda^\prime E_{\bk+\bq}}
            -
            \frac
            {\Delta^2\sin\phi_{\bk}\sin\phi_{\bk+\bq}}
            {4\lambda E_{\bk}\lambda^\prime E_{\bk+\bq}}
            \sin^2\theta.
    \end{align}
$\lambda,\lambda^\prime=\pm$ give a sign, and $a=s,c,$ and $h$ correspond to matrix elements for $s$-wave, chiral $p$-wave, and helical $p$-wave SCs, respectively.
In Eq.\ (\ref{eq_dynamic_spin_susceptibility_green_fn}), the terms multiplied by $M^{\lambda,\lambda(a)}_{\bk,\bk+\bq}$ describe the intraband transition processes, i.e., transition processes from particle to particle and from hole to hole, and the terms multiplied by $M^{\lambda,-\lambda(a)}_{\bk,\bk+\bq}$ describe the interband transition processes, i.e., transition processes from particle to hole and vice versa.
The retarded and advanced Green's functions of the quasiparticles $g^{R/A}_{\lambda,\bk}(E)$ are given by
\begin{align}
    g^R_{\lambda,\bk}(E)=\frac{1}{E-\lambda E_{\bk}+i\Gamma}, \\
    g^A_{\lambda,\bk}(E)=\frac{1}{E-\lambda E_{\bk}-i\Gamma},
\end{align}
where $\Gamma$ is a constant level broadening introduced phenomenologically.
$\Gamma$ is introduced to incorporate the intraband contribution in the calculation of the uniform spin susceptibility. The details are explained in the next section.

The sum over $\bk$ is replaced by the integral near the Fermi energy
\begin{align}
    \sum_{\bk}F(k)
        &\to D_F\int_{0}^{\infty}dE D_s(E)
            \sum_{\eta=\pm}F_{\eta}(E),    \\
    \sum_{\bk}F(k)\sin^2\phi_{\bk}
        &\to D_F\int_{0}^{\infty}dE D_s(E)
            \sum_{\eta=\pm}\frac{1}{2}F_{\eta}(E),
\end{align}
where $D_F$ is the density of states near the Fermi energy in the normal state and $D_s(E)$ is the density of states of quasiparticles
\begin{align}
    D_s(E)=\frac{|E|}{\sqrt{E^2-\Delta^2}}\theta(|E|-\Delta).
\end{align}
$F_{\eta}(E)$ means to assign $\eta\sqrt{E^2-\Delta^2}$ to $\xi$ contained in $F(k)$.

\section{Uniform spin susceptibility}

In this section, we explain three properties related to the calculation of the uniform spin susceptibility. First, the matrix element's properties are explained, which is essential to understand the qualitative difference between spin-singlet $s$-wave and spin-triplet $p$-wave SCs. Second, the reason to introduce the constant level broadening $\Gamma$. Third, the analytical expression for the uniform spin susceptibility of the $p$-wave SCs is given.

Performing the angular integral and replacing the sum over $\bk$ by the $E$ integral, the matrix elements are replaced by
\begin{align}
    &M^{\lambda,\lambda^\prime(s)}_{\bk,\bk}
    \to
    \frac{1+\lambda\lambda^\prime}{4\lambda\lambda^\prime}, \\
    &M^{\lambda,\lambda^\prime(c)}_{\bk,\bk}
    \to
    \frac
    {(1+\lambda\lambda^\prime)E^2-(1+\cos^2\theta)\Delta^2}
    {4\lambda\lambda^\prime E^2}, \\
    &M^{\lambda,\lambda^\prime(h)}_{\bk,\bk}
    \to
    \frac
    {(1+\lambda\lambda^\prime)E^2-(1+\frac{1}{2}\sin^2\theta)\Delta^2}
    {4\lambda\lambda^\prime E^2}.
\end{align}
Here, the first-order terms in $\xi_{\bk}$ are omitted because they vanish in the $E$ integral.
From the above expressions, the intraband matrix elements become finite for all SCs considered here, while the interband matrix elements vanish in the $s$-wave SC and becomes finite in the $p$-wave SCs.
The above properties of the intraband and interband matrix elements can be understood using the commutation relation between the Hamiltonian and the spin operators.
We introduce the BdG form of the spin operators $\sigma^a_{\mathrm{BdG}}$ $(a=x,y,z)$ as below
\begin{align}
    \sigma^a_{\mathrm{BdG}}=\begin{pmatrix}
        \sigma^a & 0 \\
        0 & -(\sigma^a)^{\mathrm{T}}
    \end{pmatrix}.
\end{align}
The commutation relation of $H_{\mathrm{BdG}}$ and $\sigma^a_{\mathrm{BdG}}$ is given by
\begin{align}
    \left[H_{\mathrm{BdG}},\sigma^a_{\mathrm{BdG}}\right] =   0 & :s-{\rm wave} \label{eq_comm_s}, \\
    \left[H_{\mathrm{BdG}},\sigma^a_{\mathrm{BdG}}\right]\neq 0 & :p-{\rm wave} \label{eq_comm_p}.
\end{align}
Equation (\ref{eq_comm_s}) means that both the Hamiltonian and the spin operator are diagonalized simultaneously, so that the matrix elements of the spin operator between a particle and a hole with the same wave-number vanish. This is because the $s$-wave SC is spin singlet. Therefore, the interband matrix elements vanishes in the $s$-wave SC.
In contrast, in the $p$-wave SCs, the commutation relation between the Hamiltonian and the spin operator is finite as shown in Eq.\ (\ref{eq_comm_p}), so that the matrix elements of the spin operator between a particle and a hole with the same wave-number is finite. This is because the $p$-wave SCs are spin triplet. As a result, the interband matrix elements are finite.

Here, we explain the reason to introduce the constant level broadening $\Gamma$ for $g^{R/A}_{\lambda,\bk}(E)$.
The intraband and interband transitions are schematically shown in Fig.~\ref{fig_transition}.
The quasiparticles are scattered due to the magnon emission or absorption.
The scattering process conserves the wave-number.
Consequently, in the case of the intraband transition, the transition process is forbidden when $\Gamma=0$.
In order to incorporate the intraband processes, one needs to introduce $\Gamma$, otherwise the intraband contribution vanishes, which can be directly shown by calculating Eq.\ (\ref{eq_dynamic_spin_susceptibility_green_fn}).

When $\Gamma=0$, the uniform spin susceptibility for the chiral $p$-wave SCs is given by
\begin{align}
    {\mathrm{Re}}\chi^{R}_{\mathrm{uni}}(\omega)
    =
    &2D_F\int^\infty_{\Delta}dE\frac{E}{\sqrt{E^2-\Delta^2}}
    \frac{(1+\cos^2\theta)\Delta^2}{4E^2}
    (f(E)-f(-E))\left(\frac{1}{2E+\hbar\omega}+\frac{1}{2E-\hbar\omega}\right),
\end{align}
and
\begin{align}
    {\mathrm{Im}}\chi^{R}_{\mathrm{uni}}(\omega)
    =
    &2\pi D_F\frac{|\hbar\omega/2|}{\sqrt{(\hbar\omega/2)^2-\Delta^2}}
    \frac{(1+\cos^2\theta)\Delta^2}{(\hbar\omega)^2}
    \left(
        f(-\hbar\omega/2)-f(\hbar\omega/2)
    \right).
\end{align}
From the above expressions, one can show that both the real part and imaginary part of the uniform spin susceptibility diverge at $\hbar\omega=2\Delta$, leading a resonance peak.
The expressions for the helical $p$-wave SC can be obtained by replacing $\cos^2\theta$ with $\frac{1}{2}\sin^2\theta$.
Therefore, $\theta$ dependence of $\chi^{R}_{\mathrm{uni}}(\omega)$ explained in the main text is obtained from the above expressions.

\begin{figure}
\begin{center}
\includegraphics[width=0.5\hsize]{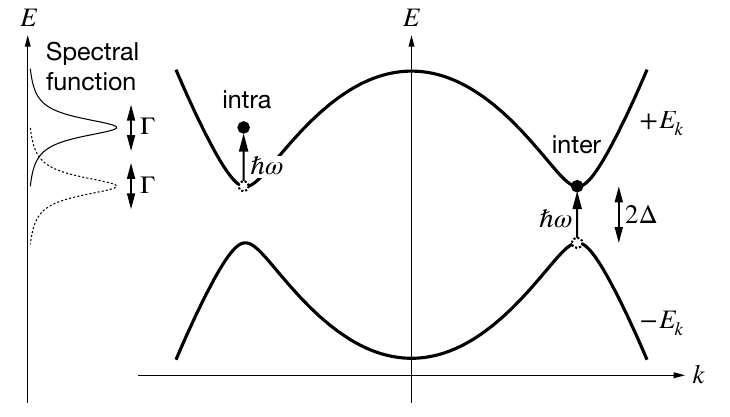}
\end{center}
\caption{
Schematic image of intraband transition and interband transitions.
The intraband transition gives contribution to the uniform spin susceptibility when the excitation energy is comparable to or smaller than the level broadening, $\hbar\omega\lesssim\Gamma$.
The interband contribution is dominant when the excitation energy is comparable to the superconucting gap, $\hbar\omega\approx2\Delta$.
}
\label{fig_transition}
\end{figure}

\section{local spin susceptibility}

Performing the angular integral and replacing the sum over $\bk,\bq$ by the $E,E^\prime$ integral, the matrix elements are replaced by
\begin{align}
    &M^{\lambda\lambda^\prime(s)}_{\bk,\bq}
    \to
    \frac{1}{4}+\frac{\Delta^2}{4\lambda E\lambda^\prime E^\prime}, \\
    &M^{\lambda\lambda^\prime(c)}_{\bk,\bq}
    \to
    \frac{1}{4}, \\
    &M^{\lambda\lambda^\prime(h)}_{\bk,\bq}
    \to
    \frac{1}{4}.
\end{align}
The matrix elements for the chiral and helical $p$-wave SCs are identical. From the above expressions, one can see that the interband contribution in the $s$-wave SC is suppressed.
Unlike the uniform spin susceptibility, the intraband contribution for the local spin susceptibility is finite even when $\Gamma=0$.
This is because the transition processes considered here leads to momentum transfer and the intraband transition is not forbidden.
Therefore, we calculate the local spin susceptibility at $\Gamma=0$.
The local spin susceptibility for the $s$-wave SC is given by
\begin{align}
    \chi^{R}_{\mathrm{loc}}(\omega)=-D_F^2
            &\int^\infty_{-\infty}dE
            \int^\infty_{-\infty}dE^\prime
            D_s(E)D_s(E^\prime)
            \left(1+\frac{\Delta^2}{EE^\prime}\right)
            \frac{f(E)-f(E^\prime)}{E-E^\prime+\hbar\omega+i0},
\end{align}
and the local spin susceptibility for the $p$-wave SCs is given by
\begin{align}
    \chi^{R}_{\mathrm{loc}}(\omega)=-D_F^2
            &\int^\infty_{-\infty}dE
            \int^\infty_{-\infty}dE^\prime
            D_s(E)D_s(E^\prime)
            \frac{f(E)-f(E^\prime)}{E-E^\prime+\hbar\omega+i0}.
\end{align}

\section{FMR modulation: Rough interface}

In this section, we show the numerical results and summarize the characteristic properties of the FMR modulation for the rough interface limit. In the following calculations, we set $J_1=0$ and assume that only $\chi^R_{\mathrm{loc}}(\omega)$ contributes to $\delta H$ and $\delta\alpha$.

Figures \ref{fig_chi_loc_sm} show (a) $\delta H$ and (b) $\delta\alpha$ for the chiral and helical $p$-wave SCs as a function of frequency and temperature.
$\delta H$ is finite in $T\to0$ and has a resonance peak at $\hbar\omega = 2\Delta$.
$\delta\alpha$ exhibits a coherence peak just below the transition temperature in the sufficiently low frequency region, where $\hbar\omega/k_{\rm B}T_c\ll1$.
$\delta\alpha$ drops abruptly at $\hbar\omega=2\Delta$.
$\delta\alpha$ is almost independent of both frequency and temperature when $\hbar\omega>2\Delta$.

Figures \ref{fig_chi_loc_sm} show (c) $\delta H$ and (d) $\delta\alpha$ for the $s$-wave SC as a function of frequency and temperature.
In the low frequency region, where $\hbar\omega/k_{\rm B}T_c\le1$, $\delta H$ at a fixed frequency decreases by about thirty percent with the decrease of the temperature, and $\delta H$ is finite in $T\to0$.
As the frequency increases, $\delta H$ is almost independent of the temperature.
$\delta\alpha$ shows a coherence peak just below the transition temperature in the sufficiently low frequency, where $\hbar\omega/k_{\rm B}T_c\ll1$.
The coherence peak in the $s$-wave SC is larger than the corresponding coherence peak in the $p$-wave SCs.
$\delta\alpha$ has a kink structure at $\hbar\omega=2\Delta$.

Note that the cutoff energy $E_c$ was introduced here to cause the integral for ${\rm Re}\chi^R_{\rm loc}(\omega)$ to converge. Although ${\rm Re}\chi^{R}_{\rm loc}(\omega)$ is approximately proportional to $E_c$, the qualitative properties explained above are independent of $E_c$.

The FMR modulation properties of the three SCs are summarized in Table \ref{table_chi_loc}.
In the case of the rough interface limit, the pairing symmetry can be detected from either the absence or the existence of the resonance peak of $\delta H$.
The pairing symmetry may also be detected from the properties of $\delta\alpha$, the height of the coherence peak, and the structure at $\hbar\omega=2\Delta$.
When compared with the resonance peak for $\delta H$, however, the properties of $\delta\alpha$ are too ambiguous to allow the pairing symmetry to be distinguished clearly.

\begin{figure}[H]
\begin{center}
\includegraphics[width=0.5\hsize]{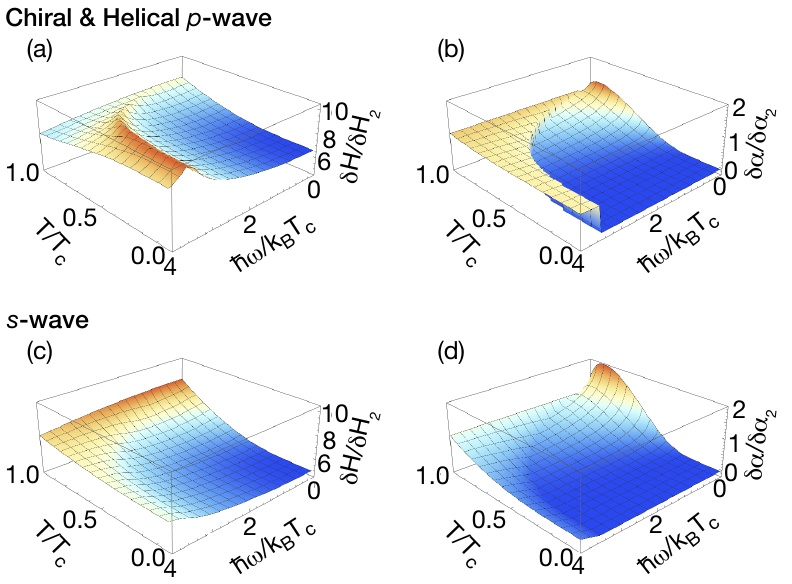}
\end{center}
\caption{
(a) The frequency shift and (b) the enhanced Gilbert damping as a function of both frequency and temperature for the $p$-wave SCs.
(c) The frequency shift and (d) the enhanced Gilbert damping as a function of both frequency and temperature for the $s$-wave SC.
The terms $\delta H_2$ and $\delta\alpha_2$ are given by $\delta H_2=-2\pi SJ_2^2l^2D^2_Fk_BT_c/(NA\gamma\hbar)$ and $\delta\alpha_2=2\pi SJ_2^2l^2D^2_F/(NA)$, where they are characteristic values in the normal state.
The cutoff energy is set to be $E_c/k_{\mathrm{B}}T_c=10$.
}
\label{fig_chi_loc_sm}
\end{figure}

\begin{table}[b]
\caption{\label{table_chi_loc}
FMR modulation properties for the rough SC/FI interface where $J_1=0$ and $J_2\neq0$.
}
\begin{ruledtabular}
\begin{tabular}{cccc}
    Pairing symmetry&
    $s$&
    Chiral &
    Helical\\
    \hline\hline
    Resonance peak of $\delta H$ &
    --&
    $\checkmark$ &
    $\checkmark$ \\
    \hline
    Coherence peak of $\delta\alpha$&
    large &
    small &
    small \\
    \hline
    $\delta\alpha$ at $\hbar\omega=2\Delta$&
    kink &
    jump &
    jump \\
\end{tabular}
\end{ruledtabular}
\end{table}

\end{widetext}


\bibliographystyle{apsrev4-1}
\bibliography{ref}

\end{document}